 \documentclass[aps,twocolumn,superscriptaddress,floatfix,nofootinbib]{revtex4-1}
\usepackage{graphicx,amsmath,amssymb,verbatim,color}
\usepackage{booktabs}
\usepackage{comment}
\usepackage{soul}
\usepackage{setspace}
\usepackage[dvipsnames]{xcolor}
\usepackage{bm}
\usepackage[utf8]{inputenc}
\usepackage[colorlinks=true,citecolor=blue,linkcolor=blue,urlcolor=blue, backref=false,pdfborder={0 0 0}]{hyperref}
\usepackage{float}
\usepackage{multirow}
\usepackage{dcolumn}% add hypertext capabilities
\usepackage{aas_macros}
\usepackage[normalem]{ulem}
\usepackage{orcidlink}

\renewcommand{\vec}[1]{\bm{#1}}

\definecolor{dgblue}{rgb}{0.118, 0.565, 1}
\definecolor{darkgreen}{rgb}{0,0.5,0}
\definecolor{mypink1}{rgb}{0.858, 0.188, 0.478}
\definecolor{red}{rgb}{1,0,0}
\definecolor{celestialblue}{rgb}{0.29, 0.59, 0.82}

\def\tpdf#1{\texorpdfstring{#1}{Lg}}

\begin{document}

\preprint{APS/123-QED}

\title{On the implications of the `cosmic calibration tension' beyond \tpdf{$H_0$} \\and the synergy between early- and late-time new physics }

\author{Vivian Poulin\orcidlink{0000-0002-9117-5257}}
 \email{vivian.poulin@umontpellier.fr}
 
\affiliation{Laboratoire univers et particules de Montpellier (LUPM), Centre national de la recherche scientifique (CNRS) et Universit\'e de Montpellier, Place Eug\`ene Bataillon, 34095 Montpellier C\'edex 05, France}

\author{Tristan L.~Smith\orcidlink{0000-0003-2685-5405}}
\email{tsmith2@swarthmore.edu}
\affiliation{Department of Physics and Astronomy, Swarthmore College, 500 College Ave., Swarthmore, PA 19081, USA}

\author{Rodrigo Calder\'on\orcidlink{0000-0002-8215-7292}}
\email{calderon@kasi.re.kr}
\affiliation{Korea Astronomy and Space Science Institute, Daejeon 34055, Republic of Korea}

\author{Th\'eo Simon\orcidlink{0000-0001-7858-6441}}
 \email{theo.simon@umontpellier.fr}
 
\affiliation{Laboratoire univers et particules de Montpellier (LUPM), Centre national de la recherche scientifique (CNRS) et Universit\'e de Montpellier, Place Eug\`ene Bataillon, 34095 Montpellier C\'edex 05, France}

\date{\today}

\begin{abstract}
The ``cosmic calibration tension'' is a $> 5\sigma$ discrepancy between the cosmological distance ladder built from baryonic acoustic oscillations (BAO) calibrated by the Planck/$\Lambda$CDM sound horizon ($r_s$) and Type Ia supernovae (SN1a) calibrated instead with the S$H_0$ES absolute magnitude, assuming the distance-duality relationship (DDR) holds. In this work, we emphasize the consequences of this tension beyond the value of the Hubble constant $H_0$, and the implications for physics beyond $\Lambda$CDM. Of utmost importance, it implies a larger physical matter density $\omega_m\equiv \Omega_m h^2$, as both the fractional matter density $\Omega_m$ and $h\equiv H_0/100$ km/s/Mpc  are well constrained from late-time data. New physics in the pre-recombination era must thus be able to decrease $r_s$ while either reducing the value of $\Omega_m$, or increasing the value of $\omega_m$.  Assuming a $\Lambda$CDM-like primordial power spectrum, this necessarily results in an increase in the clustering amplitude $\sigma_8$. 
Deviations from $\Lambda$CDM in the late-time expansion history cannot resolve the calibrator tension but can help relax the required shifts to the matter density and $\sigma_8$: it is in that sense that {\it a combination of early and late-time new physics may help alleviate the tension}. 
More precisely, models that modify the pre-recombination expansion history can accommodate the increase in $\omega_m$ without the need for additional modifications.
It is those models which only affect recombination that require additional deviations at late-times to be successful.
Hence,  the ``cosmic calibration tension'' points either to a targeted modification of the pre-recombination expansion history, or to a broader change affecting multiple cosmic epochs. 

\end{abstract}

\maketitle

\section{Introduction} 

In recent years, the prediction of the $\Lambda$-cold-dark-matter ($\Lambda$CDM) model when fit to Planck Cosmic Microwave Background (CMB) data for the expansion rate of the universe $H_0 = 67.36\pm0.54$ km/s/Mpc \cite{Planck:2018vyg} is being challenged by the ``direct'' measurement from a variety of probes of the late-time universe \cite{Verde:2023lmm}, and foremost by the cepheid-calibrated supernovae (SN1a) distance ladder by the S$H_0$ES collaboration, which measured $H_0 = 73.04\pm1.04$ km/s/Mpc \cite{Riess:2021jrx}. 
Barring the existence of unknown systematic effects (see Ref.~\cite{Riess:2021jrx,Abdalla:2022yfr} for discussion), the ``Hubble tension'' might be a hint of physics beyond $\Lambda$CDM.
\begin{figure*}
    \centering
    \includegraphics[width=0.9\textwidth] {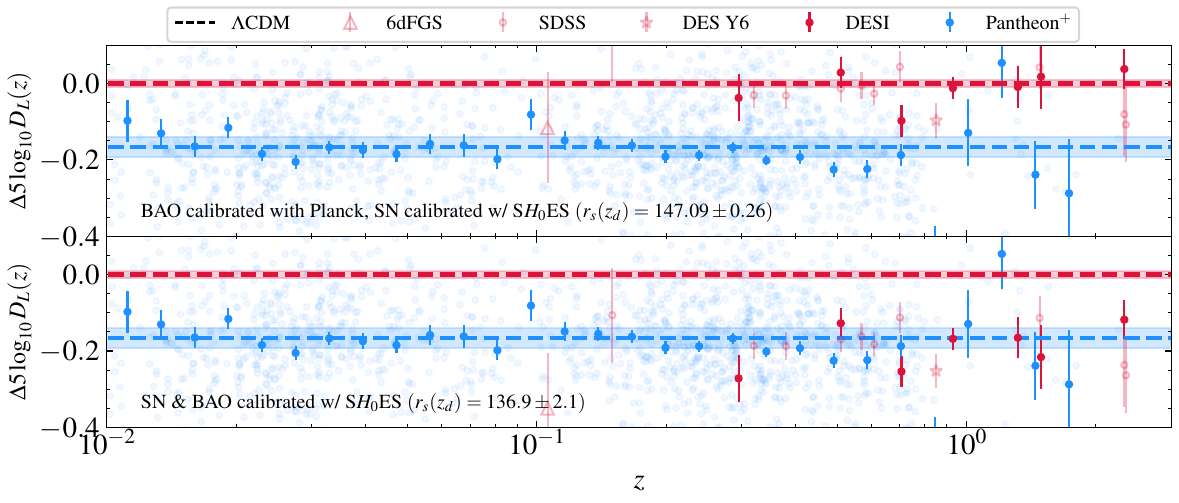}
    \caption{Luminosity distance from Pantheon$^+$, SDSS, DES and DESI data (assuming the DDR holds), normalized to the Planck/$\Lambda$CDM prediction. In the top panel, SN1a data are calibrated using the S$H_0$ES magnitude $M_b = -19.253\pm0.027$ (blue), while BAO data are calibrated assuming Planck/$\Lambda$CDM sound horizon, $r_s(z_d) = 147.09\pm0.26$ (red). Note that the solid blue measurements correspond to the binned version of the Pantheon$^+$ data, while the full dataset is shown in light blue. The overall mismatch between the two samples is the ``cosmic calibration tension''. Resolving the tension requires either changing the calibration values or breaking the DDR relation. In the bottom panel, the BAO data points are now calibrated using the value of $r_s(z_d)=136.9\pm2$ Mpc required to match S$H_0$ES. One can see that the distance diagrams are now in good agreement.
    More details on the production of this figure are provided in App.~\ref{app:hubble_diag}.}
    \label{fig:hubble_diagram}
\end{figure*}

Yet, reducing the tension between S$H_0$ES and Planck into solely a mismatch in estimates of the Hubble parameter is misleading.
Indeed, determinations of $H_0$ are ultimately tied to the calibration of the cosmological distance ladder built from SN1a and from Baryonic Acoustic Oscillation (BAO) measurements. 
When the S$H_0$ES calibration of the SN1a absolute magnitude\footnote{See Ref.~\cite{Verde:2023lmm} for a review of alternative calibrations. It is important to note that while S$H_0$ES provides the most accurate calibration, it appears to be around the median of other calibration methods (e.g. Fig.~3 in Ref.~\cite{Verde:2023lmm}). For the reminder of this paper we will focus on the S$H_0$ES cepheid calibration.} $M_b$, and the cosmic microwave background (CMB) calibration of the sound horizon at baryon drag\footnote{In detail, the sound horizon inferred from the CMB is evaluated at recombination, i.e., $r_s(z_{\rm rec})$. However, the values of $r_s(z_d)$ and $r_s(z_{\rm rec})$ are tightly correlated and relatively unaffected by the models we consider here, so that a determination of $r_s(z_d)$ effectively determines $r_s(z_{\rm rec})$ and vice versa.} $r_s(z_d)$ (extracted under the $\Lambda$CDM model) is used,\footnote{Or equivalently the Big Bang Nucleosynthesis (BBN) determination of $\omega_b$  \cite{Schoneberg:2019wmt,Schoneberg:2024ifp}.} these diagrams are mutually inconsistent in overlapping redshift range ($0.01 \lesssim z \lesssim 2$), assuming that the angular diameter distance is related to the luminosity distance as established by general relativity (which has been referred to as the Etherington ``distance-duality'' relation (DDR)) \cite{Raveri:2023zmr,Tutusaus:2023cms}. The real question is thus not whether the discrepant determinations of $H_0$ can be made to agree, but rather, can the full redshift diagram of the SN1a and BAO be made compatible?
This is explicitly illustrated in Fig.~\ref{fig:hubble_diagram}, following Refs.~\cite{Pogosian:2021mcs,Raveri:2023zmr,Tutusaus:2023cms,Bousis:2024rnb}. In the top panel, one can see that the various BAO estimates  of $D_L(z)$ \cite{DESI:2024mwx,DES:2024cme,BOSS:2016wmc,eBOSS:2020yzd}  (calibrated under Planck/$\Lambda$CDM, in red) are systematically larger than the SN1a estimates \cite{Brout:2022vxf} (calibrated with S$H_0$ES, in blue).
As those diagrams overlap in redshift, and $r_s$ is determined by pre-recombination physics, it follows that it is impossible to resolve the tension by solely adjusting the late-time expansion history. One must either change the calibration
\cite{Camarena:2021jlr,Efstathiou:2021ocp} or break the DDR relation \cite{Tutusaus:2023cms}. 

In fact, in the literature, it has been shown that when the BAO are calibrated through the SN1a they prefer a significantly lower $r_s(z_d)$ than when calibrated using the CMB or BBN measurements \cite{Bernal:2016gxb,Aylor:2018drw}. 
This is explicitly done in the bottom panel of Fig.~\ref{fig:hubble_diagram}, where the BAO data points are now calibrated using the value of $r_s$ required to match S$H_0$ES. 
One can see that the distance diagrams are now in good agreement. 

\begin{figure*}
    \centering
    \includegraphics[width=2\columnwidth]{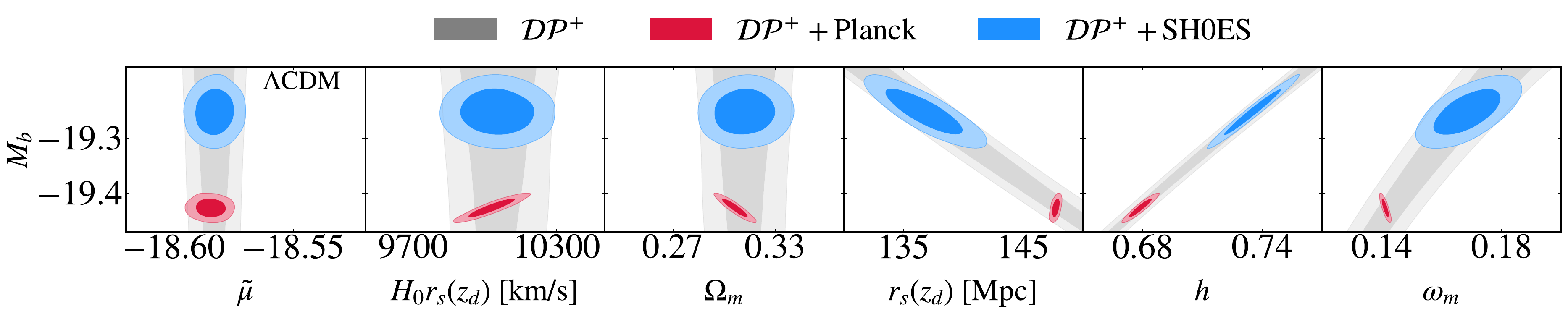}
    \caption{Posterior distributions of $M_b$ against the quantities that are well measured by uncalibrated DESI BAO and Pantheon$^+$ data (referred to as ${\cal DP}^+$) assuming the $\Lambda$CDM model holds at all times: the distance modulus at $z=0$ (up to some constant) $\tilde\mu\equiv\mu(0)=M_b-5\log_{10}(h)$, the product of $H_0r_d(z^*)$ [km/s] and the fractional matter density $\Omega_m$. When calibrating DESI, BAO, and Pantheon$^+$ with either Planck or S$H_0$ES, measurements from $H_0$  and $\omega_m$ follow. One can see that the S$H_0$ES calibration implies a small $r_s(z_d)$, a large $H_0$, and a large $\omega_m$ compared to the Planck calibration.}
    \label{fig:intro}
\end{figure*}

Reducing the value of the sound horizon $r_s(z_d)$ is, in part, what new physics must explain, and has been a main driver when building models to address this tension. 
However, it is essential to note that this is not the entire story: the SN1a calibration has several other implications, chief among them are a decrease in the age of the universe \cite{Boylan-Kolchin:2021fvy,Bernal:2021yli,Vagnozzi:2021tjv,Verde:2023lmm} and an increase in the physical matter density, $\omega_m$ \cite{Jedamzik:2020zmd,Blanchard:2022xkk,Poulin:2023lkg}.
As a result, cosmological models must affect much more than just the Hubble constant in order to provide a good fit to CMB, SN1a and BAO data under this calibration: {\it one should thus not refer to it solely as ``the Hubble tension''}. 

In this paper, we build upon previous work in the literature \cite{Bernal:2016gxb,Aylor:2018drw,Jedamzik:2020zmd,Bernal:2021yli,Boylan-Kolchin:2021fvy,Vagnozzi:2021tjv,Verde:2023lmm} by pointing out how the consequences of this ``cosmic calibration tension'' beyond $H_0$ place strict requirements on any model which attempts to provide a resolution. We then examine the dynamics of specific models with respect to these requirements. 

In a nutshell, {\it uncalibrated} BAO and SN1a give tight constraints to the shape of the expansion history over a relatively large range of redshift ($0.01 \lesssim z \lesssim 2$) and can be used to infer the `dimensionless' quantities entering in the Hubble rate, i.e. the fractional contribution to the energy density of the different species that make up a given cosmological model. {\it Calibrating} the BAO and SN1a allows us to predict the values of `dimensionful' quantities, such as physical densities and the age of the universe. These predictions can then be tested against independent measurements: we can place lower limits on the age of the universe through measurements of stellar populations in globular clusters (e.g., Refs.~\cite{Valcin:2020vav,Valcin:2021jcg,Ying:2023oie}); the physical baryon density $\omega_b$ can be inferred from measurements of the light element abundances and a model for BBN \cite{Schoneberg:2024ifp}; in the CMB, the physical matter and baryon densities affect the overall amplitude of the Sachs-Wolfe, Integrated Sachs-Wolfe effect, and the ratio of odd/even peaks (see, e.g., Ref.~\cite{Hu:1996vq}). Given that current CMB and BBN measurements are consistent with $\Lambda$CDM, any new physics which addresses the cosmic calibration tension is required to introduce specific degeneracies in order to fit all of these cosmological data sets.

Concretely, Fig.~\ref{fig:intro} demonstrates the consequences of the two calibrations under the assumption of $\Lambda$CDM. 
Uncalibrated BAO and SN1a (in gray) provide tight constraints on three quantities: the product of $H_0r_s(z_d)$, $L_b H_0^2$ (where $L_b$ is the intrinsic SN1a bolometric luminosity) which we can rewrite in terms of the absolute bolometric magnitude, $M_b$, as $\tilde \mu \equiv M_b - 5\log_{10} h$, and the fractional matter density $\Omega_m$. Extracting constraints on $h$, $r_s(z_d)$ and the physical matter density $\omega_m$ thus require calibration. As one can see, the S$H_0$ES calibration (in blue) implies a smaller $r_s(z_d)$, a larger $H_0$ and a larger $\omega_m$ compared to the Planck calibration (in red). 

All models which address the cosmic calibration tension must reduce the value of the $r_s(z_d)$ inferred from the CMB. There are two main ways the most successful models achieve this: either they propose changes to recombination so as to increase the redshift of photon decoupling or they introduce additional energy density to the pre-recombination universe. However, Fig.~\ref{fig:intro} shows that any model which attempts to {\it solely} change the pre-recombination era and fit all of the data must approximately move along the uncalibrated (gray) contours, and thus also accomodate the larger $\omega_m$: this is not an implication of the new physics introduced, but an implication of the constraints to the late-time expansion history. 
Models which introduce additional energy density have the property that they do indeed follow those degeneracy lines, whereas those which affect the redshift of recombination do not. Instead, one must introduce modifications to both the late and pre-recombination physics in order for those models to fit all of the data, as suggested in Ref.~\cite{Sekiguchi:2020teg,Vagnozzi:2023nrq,Toda:2024ncp}.  
This suggests that the ``cosmic calibration tension'' points either to a targeted modification of the pre-recombination expansion history, or to a broader change affecting multiple cosmic epochs. 

In addition, we show that within $\Lambda$CDM the ``S$H_0$ES calibrated cosmology'' predicts a significant \emph{increase} in the clustering of matter (quantified by the weak-lensing parameter $S_8$ \cite{KiDS:2020suj,DES:2021bvc, DES:2022urg,HSC:2018mrq}), and models which address these tensions in fact \emph{reduce} $S_8$.\footnote{These conclusions assume the amplitude of the primordial power spectrum is close to its value determined from Planck under  $\Lambda$CDM.} Finally, we confirm results that a measurement of the age of the universe is robust to the introduction of new physics involved in the tension and offers a promising way to independently establish (or challenge) the ``S$H_0$ES calibrated cosmology''~\cite{Boylan-Kolchin:2021fvy,Bernal:2021yli,Vagnozzi:2021tjv,Verde:2023lmm}. 

Our paper is structured as follows. In Sec.~\ref{sec:implications}, we present the implications of S$H_0$ES beyond $H_0$, discuss the requirements that must be fulfilled by a new physics model attempting to resolve the tension and the interplay between early- and late-time models. We take as examples\footnote{We stress that these `toy-models' are chosen because they are simple to implement and capture interesting phenomenological effects, rather than for their theoretical motivation.} of early modification the varying electron mass ($m_e$) model \cite{Hart:2019dxi} and the axion-like early dark energy (EDE) model \cite{Poulin:2018cxd,Smith:2019ihp}. We combine these models with the late-time modification modeled by a varying equation of state of dark energy and a curved universe. In Sec.~\ref{sec:beyondH0}, we discuss additional implications of the tension and the experimental target for surveys to test models. We eventually conclude in Sec.~\ref{sec:concl}. More details about analyses setup, model priors and datasets are provided in Apps.~\ref{app:hubble_diag} and \ref{app:MCMC}. Tables with all results are provided in App.~\ref{app:tables}. A comparison of our results with the reconstruction of the late-time dynamics with Gaussian processes is provided in App.~\ref{app:GP}. Finally, App.~\ref{app:WZDR_PMF} provide results for alternative exotic models for comparison, focusing on primordial magnetic fields (PMF) \cite{Jedamzik:2020zmd} and the Wess-Zumino Dark Radiation (WZDR) model \cite{Aloni:2021eaq}.

\section{Implications of the \tpdf{S$H_0$ES} calibration beyond \tpdf{$H_0$}} 
\label{sec:implications}
\subsection{Implications under \tpdf{$\Lambda$}CDM }
\label{sec:LCDM}

Galaxy and Lyman-$\alpha$ surveys like the Sloan Digital Sky Survey (SDSS) \cite{eBOSS:2020yzd} and the Dark Energy Spectroscopic Instrument (DESI) \cite{DESI:2024mwx} provide exquisite measurements of the transverse and longitudinal BAO, that once calibrated using an estimate of the sound horizon at baryon drag, $r_s(z_d)$, can be turned into measurements of $H(z)$ and $d_A(z)$ at $z\sim [0.1-2]$ depending on the exact probe considered. SN1a data measure apparent magnitudes $m(z)$ which, once calibrated through a measurement of their absolute magnitude $M_b$, can be turned into measurements of the  distance modulus $\mu(z)=m(z)-M_b=5\log_{10}(d_L(z)/10{\rm Mpc})$ over a redshift range which overlaps with the BAO, $z\sim[0.01,2]$.

The basic idea at the heart of the analysis is that the DDR relation between (comoving) angular and luminosity distance \cite{doi:10.1080/14786443309462220}
\begin{equation}
    d_A(z)=\frac{d_L(z)}{(1+z)^2} \,,
\end{equation}
can be used to directly compare BAO and SN1a measurements. 
This is done in Fig.~\ref{fig:hubble_diagram}, where we show the luminosity distance from Pantheon$^+$ and DESI data (assuming the DDR holds), normalized to the Planck/$\Lambda$CDM prediction. 
We also show the SDSS and DES Y6 data \cite{DES:2024cme} for comparison. 
SN1a data are calibrated using the S$H_0$ES magnitude $M_b = -19.253\pm0.027$ (blue), while BAO data are calibrated assuming Planck/$\Lambda$CDM sound horizon, $r_s(z_d) = 147.09 \pm0.26\ {\rm Mpc}$ (red).  More details about this procedure are provided in App.~\ref{app:hubble_diag}. The overall mismatch between the two samples is the ``cosmic calibration tension''. 
Interestingly, DESI and DES Y6 indicate a $\sim 2\sigma$ downward fluctuation around $z\sim 0.7-0.85$ which is closer to the SN1a calibration. Even with this, it is clear that `by eye', apart from those two points, the SN1a and BAO data are mutually inconsistent at high significance. The most constraining data in that context are in the range $z\sim0.3-0.7$, where SDSS still dominates the constraining power\footnote{In the rest of the paper, we limit our analyses to DESI data. We have checked that including SDSS does not alter our main conclusions, but can impact some of the specific numbers we provide, in particular regarding estimates of the tension level.}. 
Resolving the tension requires either a change to the calibration values or a deviation from the DDR relation.
In fact, instead of assuming that both calibrations are known, one can use either
the S$H_0$ES calibration or the Planck calibration to predict the value of the other calibrator. The ``$H_0$ tension'' can thus be interpreted as a ``calibrator tension'' as done in past literature \cite{Bernal:2016gxb,Aylor:2018drw,Camarena:2021jlr,Efstathiou:2021ocp}.
In the following, we analyze the expansion histories resulting from either calibration, in order to show that the mismatch has important implications beyond the value of $H_0$, and can help guide model building.

\begin{figure*}
    \centering
    \includegraphics[width=2\columnwidth]{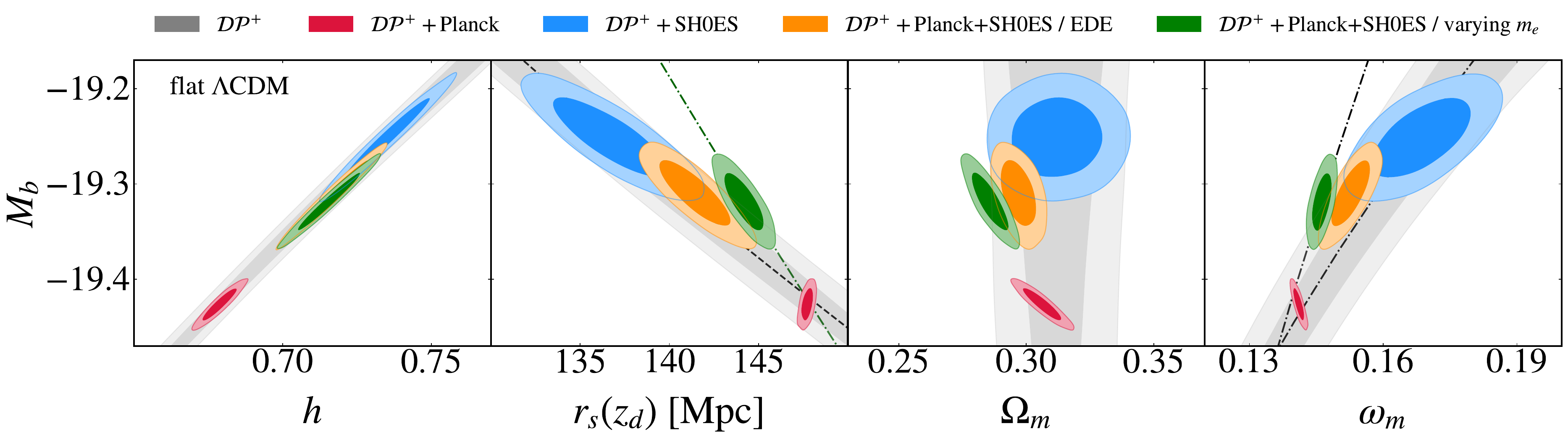}
    \includegraphics[width=2\columnwidth]{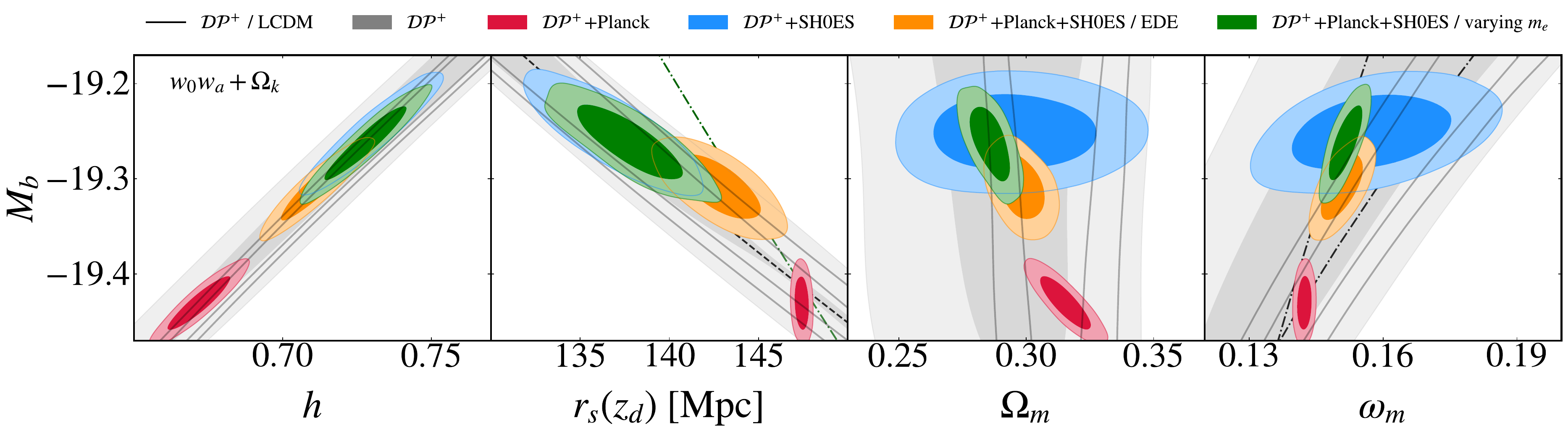}
    \caption{{\it Top panel:} Posterior distributions of $M_b$ vs $\{h,r_s(z_d),\Omega_m,\omega_m\}$ when calibrating DESI BAO and Pantheon$^+$ data (referred to as ${\cal DP}^+$) with either S$H_0$ES or Planck assuming the $\Lambda$CDM model holds at all times. We also show the posteriors in the EDE and varying $m_e$ models reconstructed from a fit to ${\cal DP}^+$+Planck+S$H_0$ES, assuming $\Lambda$CDM at late-times. EDE follows the `correct' $h \propto 1/r_s(z_d)$ relation (black dash-dotted line), while varying $m_e$ follows a $h \propto (1/r_s(z_d))^2$ relation (green dashed line). We also show the degeneracy direction  $h =\sqrt{\omega_m/\Omega_m}$ on the last panel.
    {\it Bottom panel:} Same as top panel, now  allowing for freedom in the behavior of Dark Energy and curvature at late-times. The varying $m_e$ model can now accommodate the `correct' $h \propto 1/r_s(z_d)$ relation. }
    \label{fig:LCDM_EDE_me}
\end{figure*}

We perform an initial Bayesian analysis assuming $\Lambda$CDM holds at all times for the compilation of Pantheon$^+$ SN1a data \cite{Brout:2022vxf} and DESI BAO data \cite{DESI:2024mwx,DESI:2024uvr}. More details on the analyses setup is given in App.~\ref{app:MCMC}. In the top panel of Fig.~\ref{fig:LCDM_EDE_me} we show the reconstructed posteriors of $M_b$ and its correlation with the expansion rate $h$, the sound horizon calibration $r_s(z_d)$, and the fractional and physical matter densities, $\{\Omega_m,\omega_m\}$. Gray posteriors (dataset ${\cal DP}^+$) combine uncalibrated DESI BAO and Pantheon$^+$: these represent the degeneracy directions that must be respected if the late-time dynamics is set to flat-$\Lambda$CDM. The red posteriors also include Planck NPIPE data \cite{Planck:2020olo,Rosenberg:2022sdy} (${\cal DP}^+$+Planck), and the blue posteriors include instead the S$H_0$ES calibration (${\cal DP}^+$+S$H_0$ES). 

Fig.~\ref{fig:LCDM_EDE_me} shows that the S$H_0$ES calibration leads to a larger $H_0 = 73.47\pm0.96$ km/s/Mpc than the Planck calibration $H_0= 67.88\pm0.37$ km/s/Mpc. Moreover, adjusting BAO data while calibrating with S$H_0$ES require a significantly lower sound horizon $r_s(z_d)\simeq136.9\pm2.2$ Mpc, compared to the Planck calibration prediction $r_s(z_d)=147.72\pm0.21$ Mpc, following the degeneracy $h\propto 1/r_s(z_d)$, as expected from the requirement of keeping the angular BAO scale fix.
Conversely, the predicted SN1a absolute magnitude when calibrating Pantheon$^+$ SN1a with Planck is significantly lower than that directly measured by S$H_0$ES. Models that {\it only} change $H_0$ but leave the BAO calibration unchanged are thus excluded. This is the core reason why early universe solutions are invoked to resolve the tension. 

However, there is another conclusion one can draw from this exercise. One can see that, as the {\it fractional} matter density $\Omega_m$ is well constrained from the shape of the late-time expansion history, and is in good agreement between datasets,~\footnote{Replacing Pantheon$^+$ by DESY5 \cite{DES:2024tys} or Union3 \cite{Rubin:2023ovl} would push $\Omega_m$ (and $\omega_m$) to larger values, making the problem stressed in this paper even stronger. The small tension with DESI would not affect our main conclusions. More work is needed to understand the source of differences in the matter density estimates from the various SN1a data compilations.} the additional constraints from either calibrator fixes the {\it absolute} matter density $\omega_m \equiv \Omega_m h^2$, where $H_0 \equiv h 100\ {\rm km/s/Mpc}$. Importantly, as the S$H_0$ES calibration leads to a larger $H_0$, it implies $\omega_m = 0.1681\pm0.0070$ which is significantly larger than the Planck $\Lambda$CDM calibration, $\omega_m = 0.1424\pm0.0011$. This in good agreement with previous work \cite{Jedamzik:2020zmd,Blanchard:2022xkk}.
At face value, this means that on top of reducing the sound horizon, new physics must either be able to accommodate the effects that these parameters have on the CMB and power spectra, or change the constraints to $\Omega_m$. Let us stress that the increase in  $\omega_{\rm m}$ {\it is not} a by-product of invoking early time new physics to resolve the tension. Rather, it is a requirement from the fact that the product of $\Omega_m h^2$ is well constrained once late-time data are calibrated with S$H_0$ES. New physics at early times alone {\it cannot} lead to a decrease in $\Omega_m h^2$, however, it can help compensate its effect in the CMB and matter power spectra. 

We stress that the $\Lambda$CDM (gray) contours establish the degeneracy directions that must be respected by any model that solely affects the pre-recombination history in order to provide a good fit to current measurements of the late-time expansion history. 
This will allow us to quickly identify whether a given model is capable of providing a reasonable resolution to the cosmic calibration tension.

\subsection{Role of early-time new physics} 
\label{sec:early}

As discussed in the Introduction, the basic role of early-time new physics is to reduce the size of the comoving sound horizon 
\begin{equation}
    r_s(z) = \int_z^{\infty}\frac{c_s(z') dz'}{H(z')}\,, \label{eq:rs}
\end{equation}
where $c_s(z)$ is the baryon-photon plasma sound speed, evaluated either at recombination $z_{\rm rec}\simeq 1090$ or baryon drag $z_d\simeq 1060$.
Early-time new physics can thus be split into three categories: those which increase the redshift of recombination (and baryon drag); those which increase $H(z)$; and those which decrease the plasma sound-speed.
We do not explore the latter possibility since, to the best of our knowledge, the few concrete realizations of this idea (e.g. by increasing the baryon loading effect by coupling DM to photons \cite{Boddy:2018wzy}) are ruled out as solutions to the cosmic calibrator (i.e. `$H_0$') tension. We focus instead on the case of a modified recombination or a modified expansion rate, taking as examples the varying electron mass ($m_e$) model \cite{Hart:2019dxi,Toda:2024ncp,Schoneberg:2024ynd} and the axion-like early dark energy (EDE) model \cite{Poulin:2018cxd,Smith:2019ihp} that respectively falls in the first and second category. In App.~\ref{app:PMF} and \ref{app:WZDR} we present the same analysis for the ``Primordial Magnetic Field'' \cite{Jedamzik:2020krr}  and the ``Wess-Zumino Dark Radiation''  \cite{Aloni:2021eaq} models, to show that our conclusions are not specific to the choice of model. 

We perform MCMC analyses of varying $m_e$ and EDE models using Planck, DESI BAO, Pantheon$^+$ and S$H_0$ES data. We stress that these models are known not to fully resolve the tension \cite{Hart:2019dxi,Toda:2024ncp,Schoneberg:2024ynd,Hill:2020osr,Schoneberg:2021qvd,Kamionkowski:2022pkx,Poulin:2023lkg, McDonough:2023qcu,Efstathiou:2023fbn,Khalife:2023qbu,Schoneberg:2024ynd,Seto:2024cgo,Qu:2024lpx}. Our goal is not to emphasize their respective level of success, but simply to highlight the degeneracy directions exploited by each model.
We show in Fig.~\ref{fig:LCDM_EDE_me} the same set of posteriors, comparing the results of EDE (in orange) and $m_e$ (in green) alongside those coming from calibrating DESI BAO and Pantheon$^+$ with either Planck or S$H_0$ES under $\Lambda$CDM.

\begin{figure*}
    \centering
    \includegraphics[width=2\columnwidth]{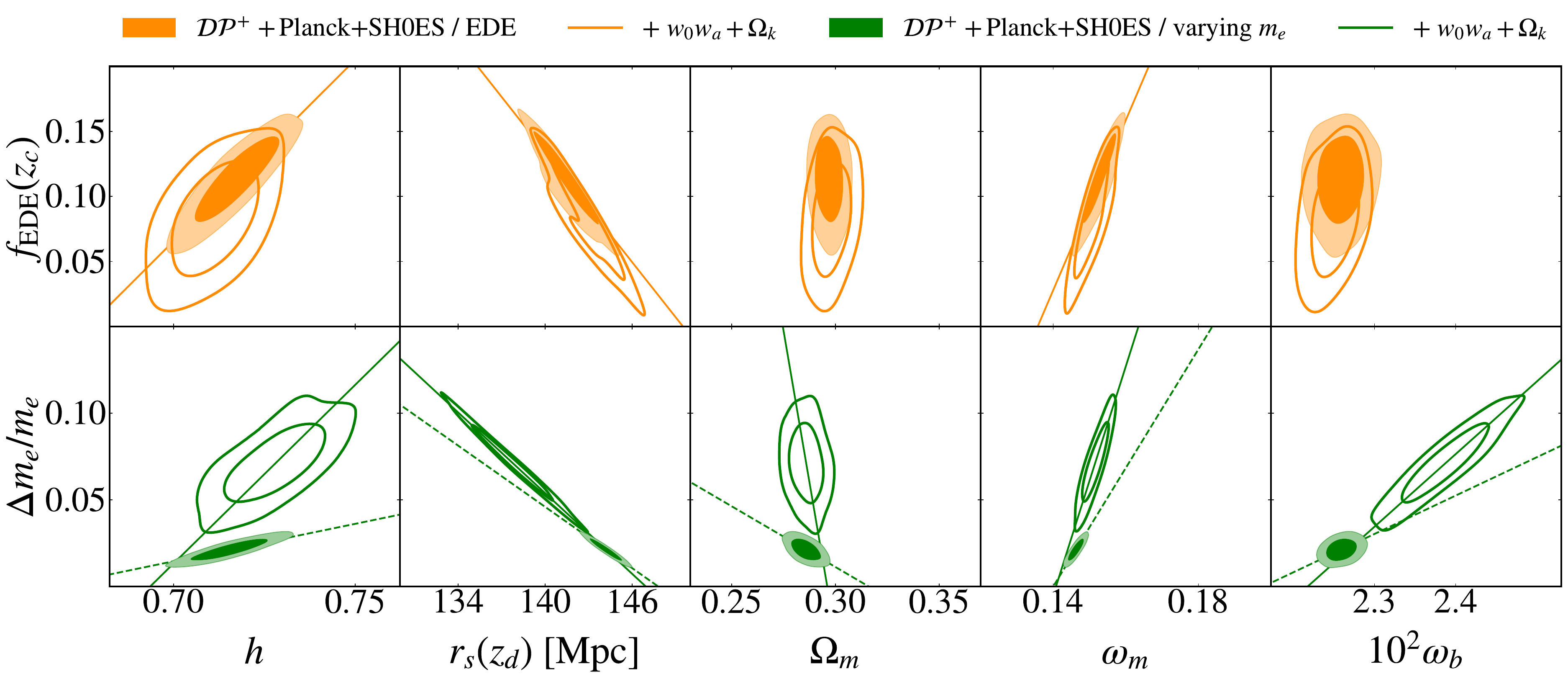}

    \caption{Posterior distributions of $f_{\rm EDE}$ and $\Delta m_e/m_e$ vs $\{h,r_s(z_d),\Omega_m,\omega_m,\omega_b\}$ reconstructed from a fit to ${\cal DP}^+$+Planck+S$H_0$ES. The full contour assumes flat $\Lambda$CDM at late-times, while the open contours leave $w_0w_a+\Omega_k$ free to vary.}
    \label{fig:correl}
\end{figure*}

One can see that both models manage to raise the inferred value of $h$, towards the high $M_b$ / low $r_s$ region. However, only the EDE model manages to respect the `correct' degeneracy directions  $r_s(z_d)\propto h^{-1} $ and $\omega_m\propto h^2$, and thus to keep $\Omega_m$ fix. On the other hand, the varying $m_e$ follows the (wrong) degeneracy line $r_s \propto h^{-2}$, and $\omega_m\propto h$, forcing $\Omega_m$ to decrease. These degeneracy directions are represented with the dashed (for EDE) and dot-dashed (for $m_e$) lines in Fig.~\ref{fig:LCDM_EDE_me}.  In addition we show in Fig.~\ref{fig:correl} the correlation between $f_{\rm EDE}$ or $m_e$ and $\{h,r_s(z_d),\Omega_m,\omega_m,\omega_b\}$. These results can be understood as follows.

For $f_{\rm EDE}(z_c)$, the requirement to keep $\theta_s$ allows us to understand the scaling with $h$: As $f_{\rm EDE}$ enters in the Hubble rate,  we have that $r_s(z_d)\propto 1/\sqrt{f_{\rm EDE}}$, and   thus $h\propto 1/r_s(z_d)\propto\sqrt{f_{\rm EDE}}$. This scaling is shown for the orange contours with an orange line in Fig.~\ref{fig:correl}, top-left panel. Hence, to keep $\Omega_m\equiv \omega_m / h^2$ fix, one needs to have that $f_{\rm EDE} \propto \omega_m$. Remarkably, this scaling leaves the fit to the CMB data unaffected, and is thus correctly found in the combined analysis (shown with solid orange line in the $f_{\rm EDE}-$vs$-\omega_m$ plane). However, this degeneracy is not perfect (i.e. valid up to any $f_{\rm EDE}$ value), preventing from alleviating the tension below the $\sim 3-3.5\sigma$ level \cite{Poulin:2023lkg,Efstathiou:2023fbn}. Fully understanding the reason for that degeneracy requires more work (see Refs.~\cite{Vagnozzi:2021gjh,Poulin:2023lkg} for discussion of the impact of EDE on the Weyl potential, the SW and the EISW term), and will hopefully lead to a more successful model of exotic pre-recombination expansion history. 

For varying $m_e$, Ref.~\cite{Sekiguchi:2020teg} finds that fixing $\theta_s$ induces a scaling $\Delta m_e/m_e \propto h^{1/3}$. This scaling is shown for the full green contour with the dashed green line in Fig.~\ref{fig:correl}, bottom-left panel. 
Keeping $\Omega_m$ fix would thus require $\Delta m_e/m_e \propto \omega_m^{1/6}$. As a compromise between the scaling required by BAO/SN1a data and that required to leave the CMB unaffected\footnote{As shown in Ref.~\cite{Hart:2019dxi,Sekiguchi:2020teg}, keeping the ratio of $a_{\rm eq}/a_*$ fixed to leave the CMB unaffected leads to a correlation $\Delta m_e/m_e \propto \omega_m$.} we find that the data favor $\Delta m_e/m_e \propto \omega_m^{1/2}$. Hence,  it is not possible to increase $h$, leave $\Omega_m$ fixed, and compensate with the required scaling of $\Delta m_e/m_e$.

In conclusion, both varying $m_e$ and EDE can help reduce $r_s$. However, EDE has the `correct' degeneracies with $\omega_m$ and $h^2$ to allow for $\Omega_m$ to stay constant. The varying $m_e$ induces a scaling that forces a smaller $\Omega_m$. We show in  App.~\ref{app:PMF} and  App.~\ref{app:WZDR} that similar conclusions are reached for other exotic recombination and exotic expansion history models. Indeed, the WZDR is also able to accommodate the `correct' scaling with $\omega_m$ and $h^2$, while PMF cannot (and follows the same degeneracy lines as the $m_e$ model). Although current models are not fully successful (i.e.~they can exploit degeneracies only for a range of $h$ and $\omega_m$ values that is too narrow to reach that measured by S$H_0$ES), respecting this multi-dimensional degeneracy with a single mechanism should be seen as promising and motivates further efforts in developing models which alter the pre-recombination expansion history. 
Alternatively, it remains to be seen whether a fully model-independent reconstruction of the recombination (and potentially also reionization) history would confirm our conclusions, e.g. following Refs.~\cite{Lee:2022gzh,Lynch:2024gmp,Lynch:2024hzh}. We note here that our results agree well with the model-independent reconstruction of Ref.~\cite{Lee:2022gzh}, that studied perturbative modifications to recombination. One would need to go further, applying for instance the methods developed in Refs.~\cite{Lynch:2024gmp,Lynch:2024hzh}.

\subsection{Role of late-time new physics}
\label{sec:late}

Given the key role played by the constraints to $\Omega_m$ provided by the BAO and SN1a data in disfavoring models that solely affects recombination, it is natural to ask whether relaxing the assumption of flat-$\Lambda$CDM at late-times can help in resolving the tension. In doing so, we want to answer two key questions: i) to what extent can a combination of early- and late-time solutions help in resolving the cosmic calibration tension? ii)  what are the consequences of the S$H_0$ES calibration of DESI BAO and Pantheon$^+$ SN1a data that are independent of the model of late-time expansion history?

\begin{figure*}
    \centering
    \includegraphics[width=2\columnwidth]{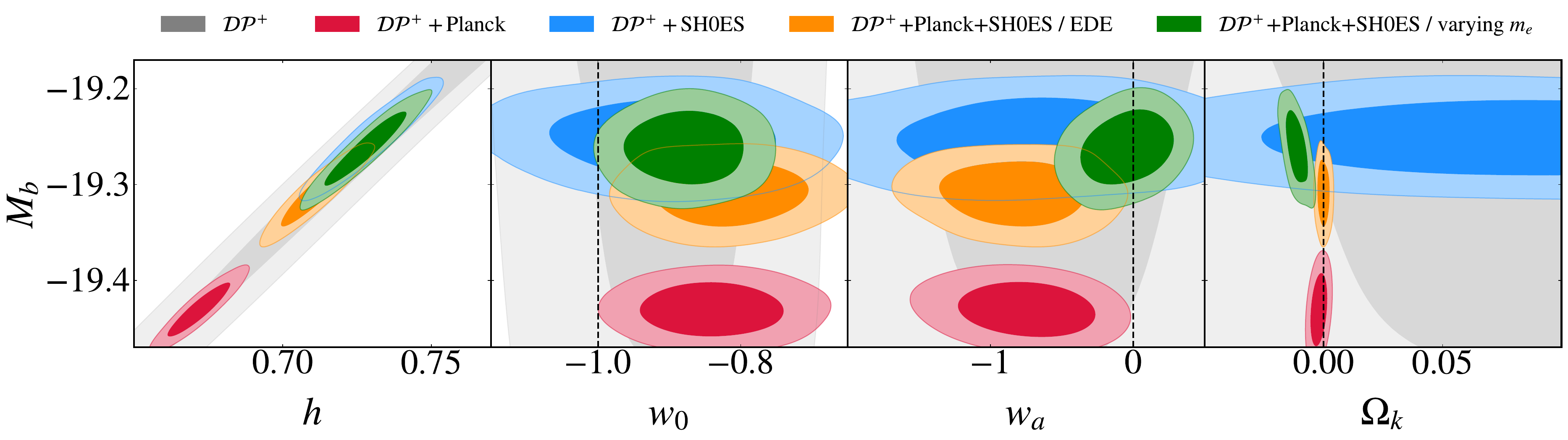}

    \caption{Posterior distributions of $M_b$ vs $\{h,w_0,w_a,\Omega_k\}$ when calibrating DESI BAO and Pantheon$^+$ data (referred to as ${\cal DP}^+$) with either S$H_0$ES or Planck. We also show the posteriors in the EDE and varying $m_e$ models reconstructed from a fit to ${\cal DP}^+$+Planck+S$H_0$ES.}
    \label{fig:CPL_Omk}
\end{figure*}

We perform the same set of analyses, but now marginalize over the behavior of Dark Energy and curvature at late-times. 
We model dark energy as a fluid (with effective sound speed $c_s^2=1$) and let its equation of state vary according to the Chevallier-Polarski-Linder (CPL) parametrization: $w(a)=w_0+w_a(1-a)$. 
In App.~\ref{app:GP} we go beyond the CPL parametrization by performing a Gaussian process reconstruction of the DE equation of state in order to show that our conclusions do not rely on this assumption (we use the CPL parametrization in the main text as it converges {\it much} faster).  Our results are presented in Fig.~\ref{fig:LCDM_EDE_me}, bottom panel. The degeneracy directions required by the combination of uncalibrated DESI BAO and Pantheon$^+$ ($\mathcal{DP^{+}}$) data are shown in gray, while the red posteriors include the Planck NPIPE data and the blue posteriors include instead the S$H_0$ES calibration. 

First, one can see as previously claimed that relaxing the late-time expansion history does not allow us to reconcile the Planck and S$H_0$ES calibration of the DESI BAO and Pantheon$^+$ SN1a. The degeneracy line between $M_b$ and $r_s(z_d)$ is indeed unaffected by the additional late-time dynamics. 
One can notice however that there is a large relaxation of the constraints to $\Omega_m$ (roughly by a factor of 2) once freedom is given in the late-time expansion history, which in turn allows for a much broader range of $\omega_m$ values. This is because, at the background level, the CPL parameterization can mimic the effect of matter, thus introducing a degeneracy that allows for lower $\Omega_m$. This indicates that, although late-time new physics cannot resolve the Hubble tension, it can contribute in relaxing the `S$H_0$ES' calibrated constraints to $\Omega_m$ and $\omega_m$. 
 In fact, if one insists on leaving $\omega_m \equiv \Omega_m h^2 = 0.1430\pm0.001$ \cite{Planck:2018vyg} invariant, the value of $h\simeq 0.73$ favored by S$H_0$ES predicts that the fractional matter density must decrease to $\Omega_m \simeq 0.268$. This is excluded by BAO and SN1a data if $\Lambda$CDM holds at late-times. However, allowing for deviations in the expansion history at late-times can help in accommodating a low $\Omega_m$, and thus potentially leave $\omega_m$ fixed. It is in that sense that {\it a combination of early and late-time new physics may help alleviate the tension}, as suggested in Ref.~\cite{Vagnozzi:2023nrq}. 
Yet, note that the standard Planck/$\Lambda$CDM value of $\omega_m$ is only reached at the $\sim 1\sigma$ limit, suggesting that a new physics model will in any case have to accommodate some level of increase in $\omega_m$.

\subsection{Combining early- and late-time new physics}

In order to demonstrate the effects of allowing more freedom in the late-time expansion history, the bottom panel of Fig.~\ref{fig:LCDM_EDE_me} shows the results of an analysis of EDE (in orange) and $m_e$ (in green) when allowing $w_0w_a+\Omega_k$ to vary simultaneously.\footnote{For a combination of EDE and varying $m_e$ with different late-time models, see Refs.~\cite{Reeves:2022aoi,Toda:2024ncp}.}
One can see that our expectation is confirmed: the varying $m_e$ model now performs significantly better despite a low $\Omega_m=0.2859\pm0.0052$. This comes however at the cost of an additional degree of freedom in the late-universe: we measure $\Omega_k = -0.0114\pm0.0031$, different from a flat universe at $\sim 3.7\sigma$, in good agreement with previous literature \cite{Sekiguchi:2020teg,Schoneberg:2021qvd}.  

As discussed in Ref.~\cite{Sekiguchi:2020teg,Schoneberg:2021qvd}, the inclusion of curvature allows an improved fit to the late-time BAO and SN1a despite a low $\Omega_m$ and a large $h$. A non-zero $\Omega_k$ allows for an additional freedom in setting $\theta_s$, leading to an altered degeneracy between $m_e$ and $h$, which now scales as $h \propto m_e^{2/3}$ (see Fig.~\ref{fig:correl}, solid line and empty contours). We also find that in order to leave the CMB invariant we must have $m_e \propto \omega_m$ (see also Ref.~\cite{Sekiguchi:2020teg}) such that the physical density increases relative to $\Lambda$CDM ($\omega_m=0.1515\pm0.0024$). However, this scaling between $m_e$, $h$, and $\omega_m$ leads to a decrease in $\Omega_m$ ($\Omega_m h^2 = \omega_m \propto m_e \propto h^{3/2} \rightarrow \Omega_m \propto h^{-1/2}$). Remarkably, this is precisely compensated for by the non-zero $\Omega_k$, allowing to fit the low$-z$ BAO and SN1a data \cite{Sekiguchi:2020teg}.
In addition, $\omega_b$ increases significantly to compensate the change in baryon drag introduced by shifting recombination (with $\omega_b \propto m_e$)  and appears to be in some tension with the BBN determination of $\omega_b$, as we discuss in more detail below.
 
We also find a preference for deviations from a cosmological constant, independent of the early-universe model, when DESI is combined with Planck (see Fig.~\ref{fig:CPL_Omk}). 
Interestingly however, the preference for non-zero $w_a$ is removed in the varying $m_e$ model, while it remains in the EDE case. This is reminiscent of the recent results of Ref.~\cite{Lynch:2024hzh} which found that an exotic recombination history can improve the fit to DESI similarly to what $w_0w_a$ achieves.  
This suggests that the apparent hint for deviation from a cosmological constant at late-times may in fact be due to i) the universe being slightly curved; ii) recombination having happened slightly differently than what our standard model predicts. This interesting possibility should be further explored in future work.

Let us finish this Section by quantifying the remaining tension with S$H_0$ES using the ``difference in the maximum a posteriori'' (DMAP) statistics \cite{Raveri:2019mxg}, 
\begin{equation}
    Q_{\rm DMAP} = \sqrt{\chi^2_{\rm min}({\rm w/~S}H_0{\rm ES}) -\chi^2_{\rm min}({\rm w/o~S}H_0{\rm ES})}\,.
\end{equation}
We find that the varying $m_e+w_0w_a+\Omega_k$ model provides an excellent fit to all datasets considered here, and the additional late-time freedom allows to significantly improve the model with virtually no tension leftover ($\sim 1.1\sigma$). 
On the other hand, the EDE model only relaxes the tension to the $\sim 3\sigma$ level irrespective of the additional late-time freedom. In fact, although we detect deviations from a cosmological constant in the EDE model, it is entirely driven by the small tension between DESI and Planck data \cite{DESI:2024mwx}, and not by the necessity to alter $\Omega_m$ in the EDE cosmology. This preference for non-canonical values of $w_0/w_a$ however has no impact on the Hubble tension.\footnote{Considering BAO together with CMB/BBN, it might naively seem possible to infer a higher value of $H_0$ by giving freedom to $w(z)$ (see for instance Fig. 10 in \cite{DESI:2024mwx}). However, we stress that the inclusion of high-$z$ SNe1a severely constrains such late-time modifications as solutions to the S$H_0$ES calibration problem.}

In agreement with previous literature \cite{Sekiguchi:2020teg}, our results hint at the possibility that the combination of exotic recombination (induced by the varying $m_e$ in this specific example) and a change in the late-time expansion history could be at play in producing the cosmic calibration tension. Yet, as these are unrelated degrees of freedom that act in conjunction to accommodate the S$H_0$ES calibration without worsening the fit to other datasets, one may argue that they are disfavored from Occam's razor point-of-view. It is fair to stress that the number of free parameters in this model remain small (only two), and that the inclusion of curvature may not be consider particularly exotic. 
Including additional data in the analysis can slightly improve constraints on the model, as we show below with the example of the BBN measurement of $\omega_b$ (see also Refs.~\cite{Khalife:2023qbu,Schoneberg:2024ynd} for up-to-date constraints).

In comparison, the EDE model can partially alleviate the tension by focusing solely on altering the pre-recombination expansion history, although the specific realization of EDE through an axion-like potential studied here does not fully resolve the tension. In particular, the shape of the energy injection, and the way perturbation in the species responsible for altering the expansion history, can affect a model's ability to resolve the tension \cite{Poulin:2023lkg,Simon:2023hlp}. Of course, it is possible that a different modification to the pre-recombination expansion history may better alleviate these tensions. For example, in Appendix \ref{app:WZDR} we explicitly show that the `Wess-Zumino Dark Radiation' model follows the same degeneracy direction as EDE and is better able to address these tensions, with a residual tension of $\sim 2.1\sigma$. 
Model-independent reconstructions of the early-universe expansion {\it \`a la} Ref.~\cite{Samsing:2012qx,Hojjati:2013oya,Moss:2021obd,Meiers:2023gft} provide a promising way to further improve over these models\footnote{It is also conceivable that an EDE scalar field triggering a variation of fundamental constants (i.e. a combination of both early-time models) could perform better than when considered independently (following e.g. Refs.~\cite{Vacher:2023gnp,Tohfa:2023zip}).}.

\section{Additional implications: targets for future experiments and models}
\label{sec:beyondH0}

\begin{figure*}
    \centering
    \includegraphics[width=2\columnwidth]{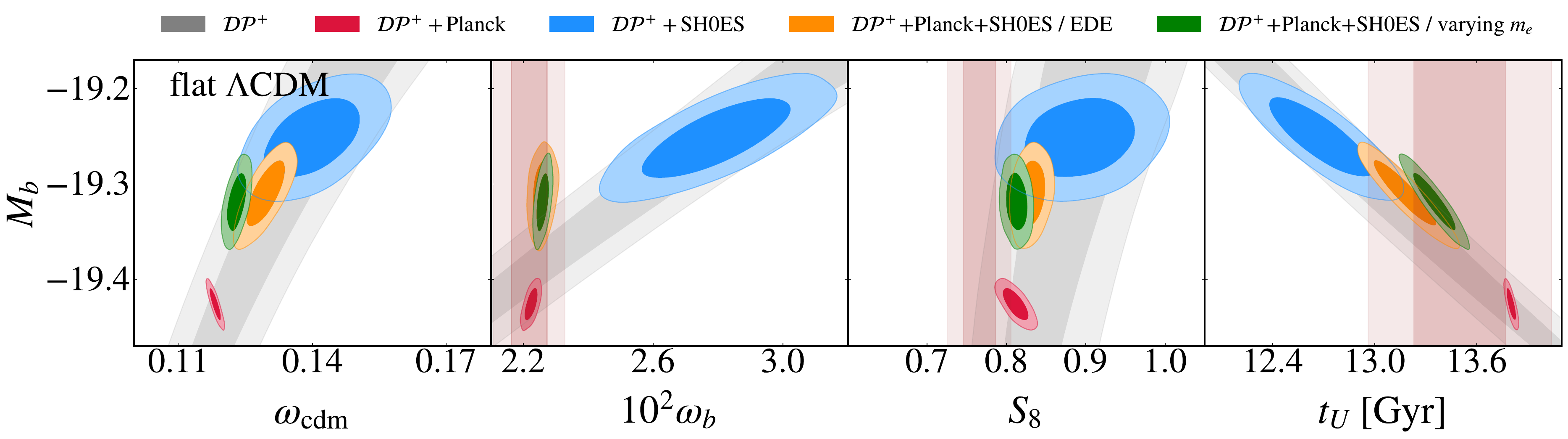}
    \includegraphics[width=2\columnwidth]{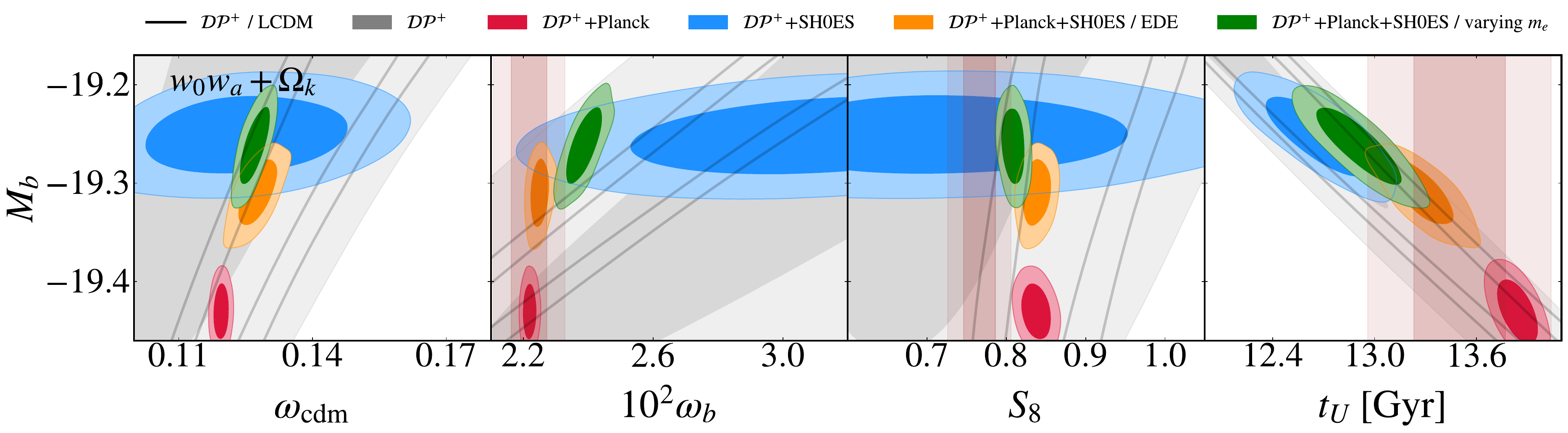}
    \caption{{\it Top panel:} posterior distributions of $M_b$ vs $\{\omega_{\rm cdm},\omega_b,S_8,t_U\}$ when calibrating DESI BAO and Pantheon$^+$ data (referred to as ${\cal DP}^+$) with either S$H_0$ES or Planck under the $\Lambda$CDM model. We also show the posteriors in the EDE and varying $m_e$ models reconstructed from a fit to ${\cal DP}^+$+Planck+S$H_0$ES. The light brown band represents the $\omega_b$ determination from BBN \cite{Schoneberg:2024ifp},  the $S_8$ determination from KiDS \cite{Heymans:2020gsg} and the $t_U$ determination from globular clusters \cite{Valcin:2020vav,Valcin:2021jcg}.
    {\it Bottom panel:} Same as the top panel, now allowing for freedom in the behavior of Dark Energy and curvature at late-times.}
    \label{fig:omb_S8_tu}
\end{figure*}

\subsection{Baryon density, growth of structure and the age of the universe}
\label{sec:omegab_s8_tu}

Though at this stage we have extensively commented on the implications of the S$H_0$ES calibration for $h$ and $\omega_m$, there are other parameters which are affected. In Fig.~\ref{fig:omb_S8_tu}, we show the 2D posterior distributions of $M_b$ vs $\{\omega_{\rm cdm},\omega_b, S_8, t_U\}$, where $\omega_{\rm cdm}$ and $\omega_b$ are the physical cold dark matter and baryon density, $S_8\equiv \sigma_8 (\Omega_m/0.3)^{0.5}$ is the clustering amplitude and $t_U$ is the age of the universe today. The top panel assumes flat-$\Lambda$CDM, while the bottom panel relaxes the late-time expansion history. We also compare the reconstructed posteriors in the EDE and $m_e$ models, with the late-time expansion given by flat-$\Lambda$CDM in the top panel, or $w_0w_a+\Omega_k$ in the bottom panel.

First,  one can see in the top panel that, under $\Lambda$CDM, the S$H_0$ES calibration (in blue) require both $\omega_{\rm cdm}$ and $\omega_b$ to increase (and not only their sum). Indeed, the larger value of $\omega_b$ is required to adjust the value of the sound horizon, while the additional increase in $\omega_{\rm cdm}$ is required to match the BAO and SN1a constraints to $\Omega_m$. However, such a large value of $\omega_b$ is in tension with other measurements. In principle, large-scale structure data alone have an independent way of measuring $\omega_b$ from the relative amplitude of the BAO peaks, which could help break this degeneracy, though this is currently not very constraining \cite{DAmico:2019fhj,Ivanov:2019pdj,Simon:2022lde,Simon:2022csv}. However, BBN measurements of $\omega_b$ \cite{Schoneberg:2024ifp} indicate significantly smaller values than what is predicted in the ``S$H_0$ES calibrated cosmology''.
This is another way of illustrating that the tension with S$H_0$ES is not solely limited to Planck CMB data: a low value of $\omega_b$ would lead to a large sound horizon, and thus to a low $H_0$ in agreement with Planck \cite{Schoneberg:2019wmt}. 
Let us note that when freedom is given at late-times (bottom panel) and the constraints on $\Omega_m$ relax, the value of $\omega_{\rm cdm}$ can be relaxed as well, as shown in the bottom panel of Fig.~\ref{fig:omb_S8_tu}. However, $\omega_b$ remains high (though error bars increase) in order to adjust the sound horizon.

In the EDE cosmology, the degeneracy between $r_s(z_d)$ and $\omega_b$ is broken ensuring not only that $r_s(z_d)$ is compatible with BAO and CMB, but also that  $\omega_b$ matches its BBN value.
In the varying $m_e$ model, however, there remains a correlation between $m_e$ and $\omega_b$ as mentioned in the previous Section. The requirement to leave the effect of baryon drag unaffected despite a larger recombination redshift requires $\omega_b \propto m_e$ (as shown in Fig.~\ref{fig:correl}, bottom right panel). Therefore, the varying $m_e+w_0w_a+\Omega_k$ model has a significantly larger $\omega_b$ value than in $\Lambda$CDM, which is in $\sim 2.5\sigma$ tension with the BBN determination \cite{Schoneberg:2024ifp}. 
In fact, performing an analysis with a BBN prior on $\omega_b$ from Ref.~\cite{Schoneberg:2024ifp}, we find that a $2.7\sigma$ tension with S$H_0$ES remains (see also Ref.~\cite{Khalife:2023qbu}). We note that this result should be taken with a grain of salt, as a proper modeling of BBN would require including the effect of the varying electron mass onto the BBN prediction (see e.g. Ref.~\cite{Schoneberg:2024ynd} that finds similar level of residual tension).

Second, the larger $\Omega_m h^2$ from the S$H_0$ES calibration implies an earlier domination of matter and therefore a larger amplitude of fluctuations in the matter power spectrum, as usually quantified by $\sigma_8$ or $S_8$. Assuming that $A_s$ and $n_s$ are well constrained to lie close to their best-fit $\Lambda$CDM values from Planck\footnote{
We stress that here we have fixed $A_s$ and $n_s$ to their best-fit $\Lambda$CDM values from Planck. While a better estimate would marginalize over the uncertainty in $A_s$ and $n_s$,  we do not anticipate our conclusions to change as those are well constrained.},  one can see from the plot that $S_8$ increases in the S$H_0$ES calibration, and is at face value at odds with direct measurements from weak lensing surveys \cite{KiDS:2020suj,DES:2021bvc, DES:2022urg}. It is also even somewhat higher than probes that are currently considered in good agreement with Planck like eROSITA cluster counts \cite{Ghirardini:2024yni} and CMB lensing \cite{ACT:2023kun}.   Let us add that the situation would worsen if we had considered the Union3 \cite{Rubin:2023ovl} or DES-Y5 SNe \cite{DES:2024tys} compilations rather than Pantheon$^+$, as those both prefer higher $\Omega_m$, and thus $\Omega_mh^2$ and $S_8$ would be even higher than the one reported in these figures. 
This demonstrates that larger $S_8$ values that typically come out of models attempting to resolve the $H_0$ tension \cite{Abdalla:2022yfr} are {\it not}  a by-product of early times new physics. They are a prediction of the S$H_0$ES calibration under $\Lambda$CDM, and new physics can actually help alleviate this issue. It may come from a change in the shape of the primordial power spectrum beyond a mere amplitude re-scaling (to respect CMB constraints); or a change in the transfer functions due to some new dynamics that may be present at early or late-times. 

 In fact, one can see in the top panel of Fig.~\ref{fig:omb_S8_tu} that early-time new physics can reduce $S_8$ compared to the naive $\Lambda$CDM prediction calibrated with S$H_0$ES. 
 However, these models do not manage to restore concordance fully with the $S_8$ determination.  
 Barring issues with non-linear modeling, baryonic feedback and intrinsic alignments modeling \cite{Amon:2022azi,Arico:2023ocu}, one needs to invoke models with exotic DM properties or modified gravity to reduce $S_8$ to the value favored by weak lensing data (see Ref.~\cite{Abdalla:2022yfr} for a review). 
 Note that allowing for freedom in the late-universe expansion (bottom panel) reduces the value of $S_8$, and in fact, when combined with an early universe model like varying $m_e$ help reducing the tension with $S_8$ to below $2\sigma$ (mainly because of the low $\Omega_m$ value).

Finally, as discussed in previous works \cite{Bernal:2021yli,Boylan-Kolchin:2021fvy,Vagnozzi:2021tjv,Verde:2023lmm}, one can clearly see that the predicted age of the universe, $t_U$ (and in fact the whole time-redshift diagram \cite{Boylan-Kolchin:2021fvy}), is different in the S$H_0$ES calibration than in the Planck cosmology. 
Under $\Lambda$CDM, S$H_0$ES predicts a value that is roughly 1 Gyr smaller than Planck. 
Interestingly,  we find that in all cases, $t_U$ cannot significantly change despite the late-time freedom introduced in the analysis. 
This stems from the fact that the integral over the inverse of the Hubble parameter entering in the computation of the age of the universe is largely weighted towards low-redshift, and dominated by the value of $H_0$. 
Increasing $H_0$ will inevitably reduce $t_U$, and the dynamics at higher redshift within the range covered by BAO and SN1a data are too constrained to allow for any significant change. 
Introducing additional new physics at early times only contribute minute change to the age of the universe.\footnote{We stress that the fact that $t_U$ in the EDE and varying $m_e$ models seems larger than the naive S$H_0$ES prediction is simply because these models do not reach the S$H_0$ES-inferred value of $H_0$.} This indicates that $t_U$ is a robust prediction of the S$H_0$ES calibration that observers should target as an independent way of testing the Hubble tension.

\subsection{Disentangling models within the CMB: role of the damping scale and the primordial tilt}
\label{sec:damping}

\begin{figure*}
    \centering
    \includegraphics[width=2\columnwidth]{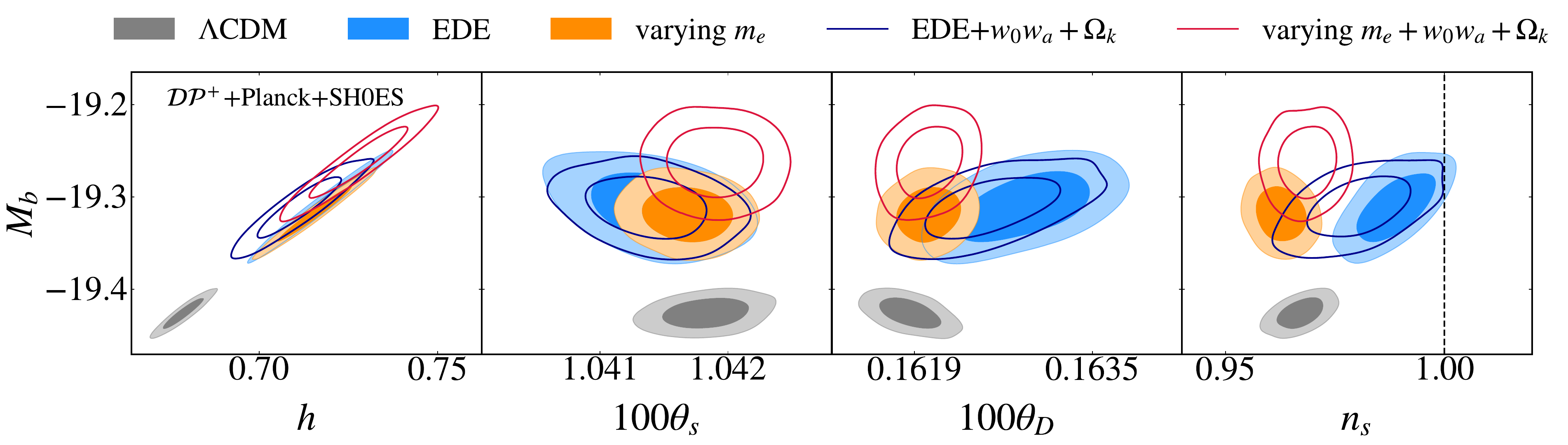}
    \caption{Posterior distributions of $M_b$ vs $\{h,100\theta_s,\theta_D,n_s\}$ in the EDE and $m_e$ models fit to Planck, DESI, Pantheon$^+$ and S$H_0$ES data. The full contours assume flat-$\Lambda$CDM at late-times, while the empty ones also leave the late-time expansion history free to vary. We also show the $\Lambda$CDM posteriors from a fit to Planck+DESI+Pantheon$^+$ for comparison. }
    \label{fig:thetaD_ns}
\end{figure*}

It has been noted in the literature \cite{Poulin:2018cxd,Takahashi:2021bti,Cruz:2022oqk,Simon:2022adh} that models like EDE prefer a larger primordial tilt $n_s$, compatible with $n_s \sim 1$, which has interesting implications for our understanding of inflation \cite{Takahashi:2021bti,Cruz:2022oqk,Simon:2022adh} (at least the robustness of some conclusions that were drawn regarding preference or exclusions of some specific models due to $n_s \sim 0.96$). This is illustrated in Fig.~\ref{fig:thetaD_ns}, where we show the posterior distributions of $M_b$ vs $\{h,100\theta_s,\theta_D,n_s\}$ in the EDE and $m_e$ models fit to Plank+DESI+Pantheon$^+$+SH0ES. 
The full contours also assume flat-$\Lambda$CDM at late-times, while the empty ones leave the late-time expansion history free to vary. 
As reviewed in Ref.~\cite{Poulin:2023lkg}, a larger $n_s$ is a by-product of the fact that while the angular size of the sound horizon $\theta_s$ can be kept fixed, the angular size of the damping scale $\theta_D$ cannot. 
It follows from the fact that the angular scale $\theta_D$ corresponding to damping roughly depends on 
\begin{equation}
    \theta_D(z_*) \sim \frac{H_0}{\sqrt{\dot \tau(z_*) H(z_*)}},
    \label{eq:theta_D}
\end{equation}
where $ \dot \tau(z_*)=n_e(z_*) x_e(z_*) \sigma_T/(1+z_*)$, $n_e$ is the electron density, $x_e$ is the ionization fraction, and $\sigma_T$ is the Thomson cross section. On the other hand, one has the approximate scaling 
\begin{equation}
    \theta_s(z_*) \sim \frac{H_0 c_s(z_*)}{ H(z_*)}\,.
    \label{eq:theta_s}
\end{equation}
If the indirect $H_0$ increases by a fraction $f$, with $\theta_s(z_*)$ fixed, $\theta_D(z_*)$ increases by $f^{1/2}$, leading to a suppression of power. 
As noticed in the past, this can be compensated for by increasing $n_s$, which in the EDE cosmology is compatible with $n_s = 1$, but similar results are obtained in the WZDR cosmology (see App.~\ref{app:WZDR}). Note that allowing for freedom in the expansion history at late-times broadens the $n_s$ posteriors toward the $\Lambda$CDM expectation, while slightly shifting $\theta_D$ as well. 

However, the varying $m_e$ cosmology has a different impact on $\theta_D(z_*)$ than EDE. Indeed, it is striking that both $\theta_s$ and $\theta_D$ can be kept fixed, and $n_s$ is left unaffected.
This is because of the additional scaling of $\sigma_T\propto m_e^{-2}$, which precisely cancels out the shifts in $h$ and in recombination redshift $z_*$ in the damping scale \cite{Hart:2019dxi,Sekiguchi:2020teg}. For comparison, we provide the results of the same analysis in the primordial magnetic field (PMF) model in App.~\ref{app:PMF}. 
It is striking that the PMF model does not perform as well despite opening the late-time expansion history. 
A key difference is precisely that the PMF model does not impact $\sigma_T$, such that $\theta_D$ is significantly shifted, contrary to the varying $m_e$ model. Ref.~\cite{Hart:2019dxi} finds similarly that varying the fine structure constant $\alpha_s$, which enters with a different scaling in $\sigma_T$, is also much less efficient at alleviating the cosmic calibration tension.
We defer to future work to investigate further reasons the differences with the PMF model. However we note here that the `clumping model' of Ref.~\cite{Jedamzik:2020krr} that we use here to study PMF is likely too approximate to fully capture the effects of PMF, and it would be worth investigating further whether a more accurate modeling could alter our conclusions.

\begin{figure}
    \centering
    \includegraphics[width=1\columnwidth]{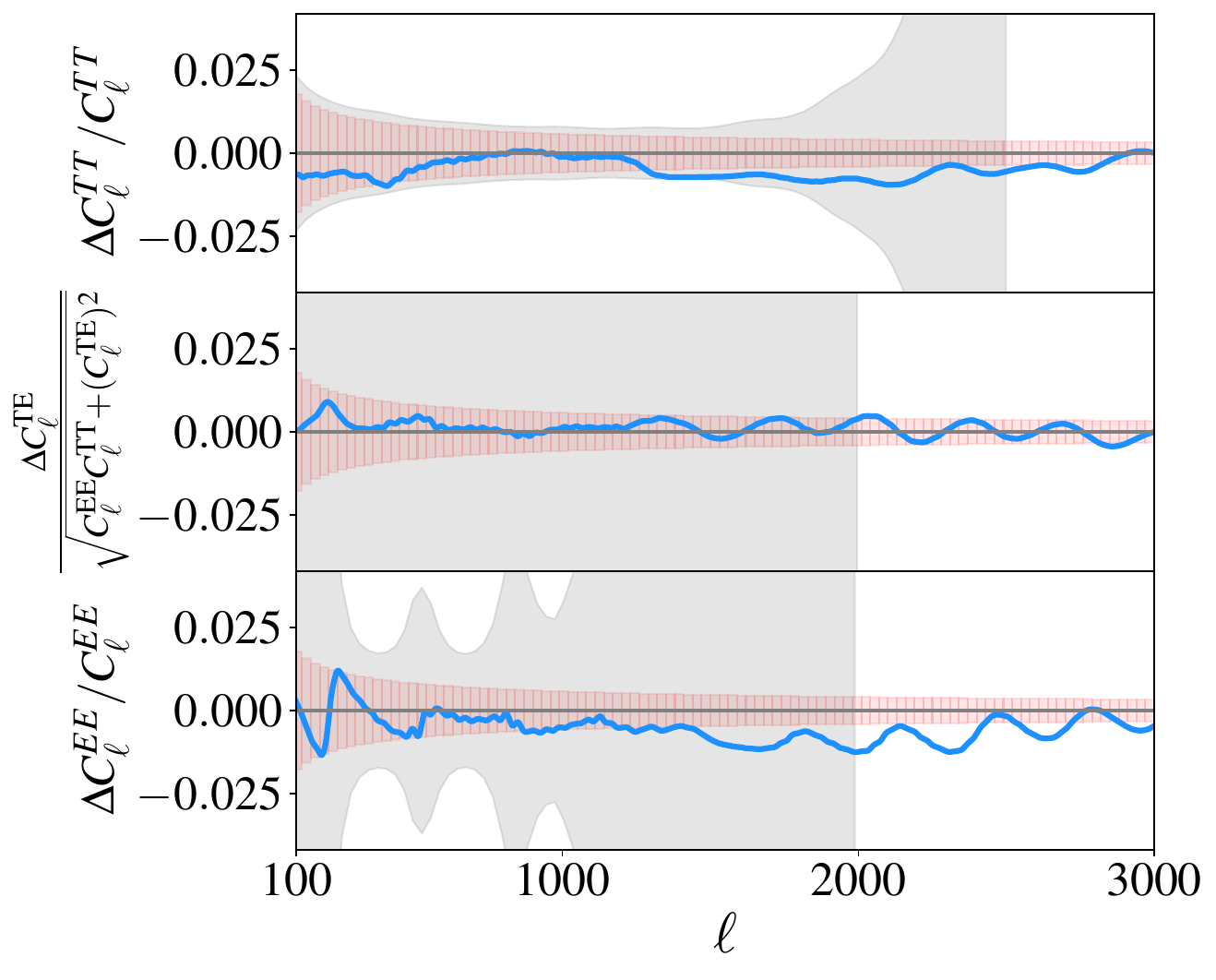}
    \caption{Residuals of the TT, TE and EE power spectra between the bestfit $m_e+w_0w_a+\Omega_k$ and EDE models. EDE is taken as reference. The gray band represent the binned error bars from Planck data release PR3, while the red boxes represent the cosmic variance with bin $\Delta \ell = 30$. }
    \label{fig:residuals}
\end{figure}

Finally, let us add that there is a way to break the degeneracy between EDE and varying $m_e$+$\Omega_k$ solely from the CMB.
Indeed, the diffusion damping scale and the primordial tilt do not affect the CMB in the exact same way, since $C_\ell^{TT} \propto \exp(-(\ell/\ell_D)^2)\times (\ell/\ell_*)^{n_s-1}$, where  $\ell_D=\pi/\theta_D\simeq 2000$ while $\ell_*$ is the angular pivot scale $\ell_*\simeq k_*/D_A(z_{\rm rec})\simeq500$. This means that the linear correlation that is exploited by the data in the EDE model is only true close to $\ell_D$. 
Hence, accurate measurements of the damping tail at intermediate  ($\ell \ll 2000$) or high-multipoles ($\ell \gtrsim 2000$) will help break that degeneracy and disentangle between models (see Ref.~\cite{Smith:2023oop}). This is illustrated in Fig.~\ref{fig:residuals}, where we show the residuals of the TT,TE and EE power spectra between the best fit $m_e+\Omega_k+w_0wa$ and EDE models, with EDE taken as reference, at $\ell\in [100,3000]$. The gray band represent the binned error bars from Planck data release PR3,\footnote{These are available at \url{https://pla.esac.esa.int/\#cosmology}.  The error bars of Planck TE extend beyond the axis limits.} while the red boxes represent the cosmic variance with bin $\Delta \ell = 30$. It is clear that a cosmic variance limited measurement  at $\ell\sim1500-2500$ in TT and EE can differentiate between both models. Near future measurements of the CMB at these multipoles (i.e., Ref.~\cite{SimonsObservatory:2018koc}) will help to further distinguish between the various solutions to the Hubble tension.
In addition, measurements of CMB spectral distortions can provide strong constraints to the primordial power spectrum, through the effect of small-scale damping, and thus help disentangle between models \cite{Schoneberg:2020nyg,
Lucca:2020fgp}. In fact, as the era of recombination itself is different in those models, the expected recombination spectrum is affected and future spectral distortion measurements will provide an important tool in the search for a solution to this tension \cite{Hart:2022agu,Lucca:2023cdl}.

\section{Conclusions} 
\label{sec:concl}

The Hubble tension is commonly phrased as a mismatch between estimates of the Hubble parameter between the CMB (assuming $\Lambda$CDM) and cepheid-calibrated SN1a. 
However, at its core, it is a mismatch between distance calibrators: the calibration of the distance ladder (that is built from the combination of SN1a and BAO assuming that the DDR holds) when using either CMB or cepheids is in disagreement.  In this work, we have emphasized that the ``cosmic calibrator tension'' has important consequences beyond the mere value of $H_0$, and that one should not reduce this to a ``$H_0$ tension'' at the risk of studying models irrelevant to the true problem. By calibrating the distance ladder using either the CMB or the S$H_0$ES SN1a magnitude, and under different assumptions about the late-time expansion history, we have identified shifts in cosmological parameters that new physics must be able to accomodate (or avoid) to solve the ``cosmic calibration tension''. Our results can be summarized as follows. 

Fig.~\ref{fig:LCDM_EDE_me} illustrates the challenges that new physics is facing in resolving the tension:

\begin{itemize}
    
\item The S$H_0$ES calibration of the SN1a magnitude requires the comoving sound horizon calibration of the BAO to decrease by $\sim 10$ Mpc, in order for the measurements of angular and luminosity distance to match between the two probes. This is true irrespective of the freedom that is given to the late-time expansion history.
\item Consistently calibrating BAO and SN1a according to the S$H_0$ES calibration leads to a large $H_0$, as expected, and the value of $H_0$ is only weakly affected by the freedom given at late-time.
    \item  If one assumes that $\Lambda$CDM holds at late-times (as commonly done with ``early universe'' models), it is necessary to accommodate the unavoidable increase in $\omega_{\rm m}$ that is a by-product of both $\Omega_m$ and $h^2$ being well constrained by late-time data. This can be naturally done in models that change the pre-recombination expansion history, like EDE or extra relativistic species. On the other hand, models that solely affect the physics of recombination, like the varying electron mass and PMF models studied here, do not follow the `correct' degeneracy with $h$ and $\omega_m$, leading to a low $\Omega_m$, and are thus disfavored in a combined analysis.
    \item Relaxing the late-time expansion history (as done here through $w_0w_a+\Omega_k$ and Gaussian Process in the appendix) can relax the constraints to $\Omega_m$, and help the varying $m_e$ model in addressing the tension by changing the correlation between $m_e$ and $h$, while accommodating a lower $\Omega_m$. Interestingly, in that case, the preference for deviations from a cosmological constant observed in DESI \cite{DESI:2024mwx,DESI:2024aqx,DESI:2024kob} decreases, while a closed universe is favored (Fig.~\ref{fig:CPL_Omk}). 
    This suggests an alternative interpretation of the hint of a deviation from a cosmological constant at late-times.
    On the other hand it has no impact for the EDE model, as the scaling between $f_{\rm EDE}$, $h$ and $\omega_m$ is unnaffected by the additional late-time freedom.
    \end{itemize}
    
There are additional implications of the S$H_0$ES calibration that our analysis has emphasized, shown in Fig.~\ref{fig:omb_S8_tu}:
\begin{itemize}

    \item If $\Lambda$CDM is assumed in the pre-recombination era, an increase in $\omega_b$ is required to adjust the value of the BAO sound horizon $r_s(z_{\rm drag})$. That increase in $\omega_b$ is strongly excluded by measurements of light element abundances and standard BBN (irrespective of the CMB). Early universe models can alter the correlation between $r_s(z_{\rm drag})$ and $\omega_b$, thus preventing a tension with BBN data.  While this works well in EDE and WZDR (two models that affect the pre-recombination expansion history), the varying $m_e$ model cannot prevent a residual $2.5\sigma$ tension with BBN data.
    \item Assuming the standard primordial power spectrum favored by Planck, the S$H_0$ES calibration implies an increase in $\omega_{\rm m}$ and an associated increase in the amplitude of fluctuations $S_8$ that, at face value, is in tension with probes of the growth of structures at low-redshift \cite{KiDS:2020suj,DES:2021bvc, DES:2022urg,ACT:2023kun,Ghirardini:2024yni} (and much larger than standard Planck $\Lambda$CDM cosmology). New physics thus needs to {\it decrease} $S_8$. Early-time models do, in fact, decrease $S_8$, though ultimately not enough to fully resolve the tension. In the varying $m_e$ model, allowing for freedom in the late-universe expansion reduces the tension with the weak lensing inferred value of $S_8$ to below $2\sigma$  (because of the lower $\Omega_m$).
     \item The ``S$H_0$ES calibrated cosmology'' predicts an age of the universe that is significantly lower than in  the ``Planck $\Lambda$CDM cosmology'', irrespective of the early- and late-time dynamics. This suggests that more experimental and theoretical efforts should be made in order to test this robust prediction of the S$H_0$ES measurement.
     \item Models which affect the early-universe expansion history predict a change in the damping scale $\theta_D$, resulting in a large increase in $n_s \sim 1$ (Fig.~\ref{fig:thetaD_ns}). On the other hand models affecting the recombination history {\it and} the Thomson scattering rate (like the varying $m_e$ model) can leave $\theta_D$ and $n_s$ unaffected. As shown in Fig.~\ref{fig:residuals}, more accurate small-scale CMB measurements will be able to break the $\theta_D-n_s$ degeneracy to help disentangle between models. Measurements of CMB spectral distortions, sensitive to the different $n_s$ values and recombination era, will also play an important role in distinguishing between models.

\end{itemize}

Our results suggest that the solution to the Hubble tension may either come from a localized modification to the pre-recombination expansion history, or to a broader change affecting both recombination and the late-time expansion history. It is also important to bear in mind that the models studied here remain `toy-models', that we use for their simplicity to implement in a concrete analysis. Yet, we expect our conclusions to be fairly generic, as explicitly checked with different implementations of these ideas in Appendix. Additional theoretical insight would certainly be valuable, in order to build models that may entail phenomenological effects not contained within the models studied here. 
Looking forward, if one favors invoking a single new mechanism (though with several free parameters), it may be possible to improve the fit by using a model-independent approach ~\cite{Samsing:2012qx,Hojjati:2013oya,Moss:2021obd,Meiers:2023gft}. We note that the data we have studied favor the latter possibility, as the varying $m_e+w_0w_a+\Omega_k$ can bring Planck+DESI+Pantheon$^+$+S$H_0$ES in agreement at $1.1\sigma$, while a tension remains at $3.1\sigma$ in the axion-like EDE model, and at $2.1\sigma$ with WZDR. However the varying $m_e$ model introduces a slight tension with the BBN inferred value of $\omega_b$. A model-independent approach to modified recombination \cite{Lee:2022gzh,Lynch:2024gmp,Lynch:2024hzh} may provide an improved fit to the data, independently from the additional late-time mechanism. 
Both these paths produce distinct signatures in cosmological observables that future surveys will help disentangle, hopefully providing clarity to the potential resolutions to the cosmic calibration tension.

\textbf{Acknowledgments} -- We thank Jens Chluba, Miguel Escudero, Laura Herold, Marc Kamionkowski, Gabriel Lynch, Adam Riess, Nils Sch\"oneberg, Sunny Vagnozzi for discussion and comments on an earlier draft.  
In addition, we would like to thank all the participants and organizers of the ModIC 2024 workshop (\url{https://indico.gssi.it/event/606/}) for the very fruitful interdisciplinary exchange of ideas. VP is supported by funding from the European Research Council (ERC) under the European Union's HORIZON-ERC-2022 (grant agreement no.~101076865). VP and TS are supported by the European Union's Horizon 2020 research and innovation program under the Marie Sk{\l}odowska-Curie grant agreement no.~860881-HIDDeN. The authors acknowledge the use of computational resources from the LUPM's cloud computing infrastructure founded by Ocevu labex and France-Grilles. TLS is supported by NSF Grants No. 2009377 and No. 2308173. This work used the Strelka Computing Cluster, which is supported by the Swarthmore College Office of the Provost.

\appendix
\section{Details about the bayesian analyses and the data likelihoods}
\subsection{Hubble diagram}\label{app:hubble_diag}

In Fig.~\ref{fig:hubble_diagram}, we show the luminosity distance from Pantheon$^+$, SDSS, DES, and DESI data (assuming the DDR holds), normalized to the Planck/$\Lambda$CDM prediction. In the top panel, SN1a data are calibrated using the S$H_0$ES magnitude $M_b = -19.253\pm0.027$ (in blue), while BAO data are calibrated assuming Planck/$\Lambda$CDM sound horizon, $r_s(z_d) = 147.09 \pm0.26$ \cite{Planck:2018lbu} (in red). In the bottom panel, the BAO data points are now calibrated using the value of $r_s(z_d)=136.9\pm2$ Mpc required to match S$H_0$ES. Note that while the DESI likelihood is built using the 12 measurements reported in Table 1 of \cite{DESI:2024mwx}, in Fig.~\ref{fig:hubble_diagram} we only show the bins where the S/N is high enough to constrain the transverse ($D_M/r_d$) and radial ($D_H/r_d$) directions of the BAO feature. For the low- and high-$z$ measurements (BGS and QSO), DESI only measures the isotropic (volume-averaged) quantity $D_V/r_d$. We translate $D_V/r_d$ assuming the $\Lambda$CDM cosmology, but we have checked that assuming the CPL bestfit mode would not affect those by a significant amount. For comparison, we also show the recent photometric measurement of $D_M(z=0.85)/r_d$ by the DES collaboration \cite{DES:2024cme} along with previous measurements by SDSS \cite{BOSS:2016wmc,eBOSS:2020yzd} and the lowest redshift measurement at $z=0.106$ from the 6dFGS \cite{Beutler:2011hx}. We emphasize however that \textit{these are not} included in the analyses.

For the Pantheon$^+$ SNe1a compilation \cite{Brout:2022vxf}, we obtain the binned distance measurements by dividing the redshift range $0.01<z<2.26$ into 30 bins, logarithmically spaced in $z$. For each redshift bin $\tilde z_i$, we select those distance moduli, $\vec\mu_i=\vec{\mu}(z)\in \tilde z_i$, belonging to the $i$-th bin. The weights associated with that redshift bin are then obtained by taking the inverse of the covariance matrix
\begin{equation}
    \vec{w}_i=\vec{C}^{-1}\in \tilde z_i~,
\end{equation}
where $\vec{C}$ is the sub-covariance matrix associated with those measurements falling within the $i$-th redshift bin, $\tilde z_i$. The mean distance modulus at each bin is then computed as 
\begin{equation}
    \tilde\mu_i = \frac{\sum \vec{w}_i \cdot\vec{\mu}_i}{\sum \vec{w}_i}~,
\end{equation}
and the associated standard deviation is given by
\begin{equation}
    \tilde\sigma_i= \sqrt{1/\sum \vec{w}_i}~,
\end{equation}
where the summation is done over all measurements within a given redshift bin.
We stress however that the binning scheme is purely for plotting/pedagogical purposes and that the analysis was carried-out using the full (un-binned) Pantheon$^+$ dataset, shown as light-blue dots in Fig. \ref{fig:hubble_diagram}.

\subsection{Data analysis}\label{app:MCMC}

We use the DESI BAO data presented in Refs.~\cite{DESI:2024uvr,DESI:2024mwx,DESI:2024lzq} and the SN1a compilation from Pantheon$^+$ \cite{Brout:2022vxf}. Regarding DESI, we make use of the data presented in Tab.~1 of Ref.~\cite{DESI:2024mwx}, that are compilation of low redshift galaxies of the bright galaxy survey, luminous red galaxies, emission line galaxies, quasars, and Lyman$-\alpha$  forest quasars tracing the distribution of neutral hydrogen. Those data cover an effective redshift range of $z\sim 0.1-4.1$. 

Regarding Pantheon$^+$, we make use of their catalog of uncalibrated luminosity distance of type Ia supernovae (SNeIa) in the range ${0.01<z<2.3}$~\cite{Brout:2022vxf}.
In order to calibrate the data, we perform sets of analyses with a prior on the intrinsic magnitude $M_b=-19.253\pm 0.027$ as measured by S$H_0$ES \cite{Riess:2021jrx}. We have checked that it provides results equivalent to using the true cepheid-distance calibration of the SN1a hosts, as presented in Ref.~\cite{Riess:2021jrx}.

The CMB data come from Planck NPIPE \cite{Planck:2020olo}, an update from the Planck 2018 data \cite{Planck:2018vyg}, through the recent \texttt{CamSpec} TTTEEE likelihood \cite{Efstathiou:2019mdh,Rosenberg:2022sdy}. We anticipate that similar results would have been obtained with the alternative \texttt{Plik} \cite{Planck:2018vyg} or \texttt{Hillipop} \cite{Tristram:2023haj} likelihoods. We include the Planck18 lensing reconstruction \cite{Planck:2018lbu} and low-$\ell$ TT and EE likelihoods \cite{Planck:2019nip}. 

We run Monte Carlo Markov Chains using the Metropolis-Hastings algorithm from \texttt{MontePython-v3}\footnote{\url{https://github.com/brinckmann/montepython_public}} code~\cite{Brinckmann:2018cvx,Audren:2012wb} interfaced with \texttt{CLASS}~\cite{Lesgourgues:2011re,Blas:2011rf}. Cosmological and nuisance parameters are varied according to Choleski's ``fast'' and ``slow'' parameters decomposition \cite{Lewis:2013hha}. We consider chains to be converged with the criterion $R-1<0.05$, though most chains have $R-1<0.01$. We analyse chains and produce figures using the python package \texttt{getdist} \cite{Lewis:2019xzd}.

\subsection{Models and priors}

All our runs consider at least the baryon density $\omega_b$, the cold dark matter density $\omega_{\rm cdm}$ and the Hubble constant $H_0$. When running with Planck NPIPE data we also include the primordial amplitude $A_s$ and tilt $n_s$ as well as the optical depth to reionization $\tau_{\rm reio}$ as free parameters. We use broad flat priors on all parameters to ensure our posteriors are not influenced by the priors. We model neutrinos as two massless, one massive species with $\sum m_\nu = 0.06$ and $N_{\rm eff}=3.046$. When including Planck data, we also use the non linear prescription \texttt{hmcode} in \texttt{CLASS} \cite{Mead:2020vgs}, though this does not have a significant affect on our analysis.

When allowing for additional freedom in the late-time expansion history, we model dark energy as a fluid (with effective sound speed $c_s^2=1$) and let its equation of state vary according to the Chevallier-Polarski-Linder (CPL) parametrization: $w(a)=w_0+w_a(1-a)$. For the priors, we follow the choice of the DESI collaboration \cite{DESI:2024mwx} and set $w_0\in[-3,1]$ and $w_a\in[-3,2]$. We also perform runs with a broad flat prior on the curvature parameter $\Omega_k\in[-0.5,0.5]$. Our conclusions are unaffected by the choice of priors. 

The EDE component is modeled through the Klein-Gordon equation with a potential $V(\phi)=m^2f^2(1-\cos(\phi/f))^3$, where $\phi$ is the field value, $f$ the axion decay constant and $m$ the axion mass. This potential has been found in the literature to provide a promising toy-model to resolve the tension \cite{Poulin:2018cxd,Smith:2019ihp,Poulin:2023lkg}, though current constraints with the NPIPE data are strengthening the constraints \cite{Efstathiou:2023fbn}. We make use of the public code \texttt{AxiCLASS}.\footnote{\url{https://github.com/PoulinV/AxiCLASS}} In practice, we follow the standard procedure \cite{Smith:2019ihp} and use a shooting algorithm to run on the fractional contribution of the early dark energy $f_{\rm EDE}(z_c)$ at the critical redshift $z_c$ where the EDE field starts to roll down its potential. We impose the following priors on the model parameters: $f_{\rm EDE}(z_c)\in [0,0.3]$, $\log_{10}(z_c)\in [3,4]$, $\theta_i\in [0,3.1]$, where $\theta_i=\phi_i/f$ is the initial field value.

For the varying electron mass model, we use the standard \texttt{CLASS} implementation,\footnote{\url{https://github.com/lesgourg/class_public}} described in Ref.~\cite{Schoneberg:2021qvd}, where the electron mass is assumed to be different than the standard mass by an amount $\Delta m_e/m_e\in[0.5,1.5]$ and transitions back to the standard model value at late-times (arbitrarily set to $z=50$).

\section{Parameter tables}
In this appendix, we provide the reconstructed mean $\pm 1\sigma$  of parameters in the $\Lambda$CDM model in Tab.~\ref{tab:cosmoparam} and $w_0w_a+\Omega_k$ model in Tab.~\ref{tab:cosmoparam1},  in light of  DESI and Pantheon$^+$ data either uncalibrated, calibrated with Planck or with S$H_0$ES.
We additionally provide reconstructed mean $\pm 1\sigma$  of parameters in the EDE and varying $m_e$ models in Tab.~\ref{tab:cosmoparam2}, and in the PMF and WZDR models in Tab.~\ref{tab:cosmoparam3}, in light of the data combination Planck+DESI+Pantheon$^+$+S$H_0$ES.
\label{app:tables}

\begin{table*}[]
    \centering
    \begin{tabular}{|l|c|c|c|}
     \hline
        \multicolumn{4}{|c|}{$\Lambda$CDM}\\
        \hline\rule{0pt}{3ex}
        & $\mathcal{DP}^+$ & $\mathcal{DP}^+$+Planck & $\mathcal{DP}^+$+S$H_0$ES\\

\hline
$\Omega_m$ & $0.312\pm 0.012$  & $0.3063\pm 0.0049$& $ 0.312\pm 0.012$\\
$H_0r_d(z^*)~$[km/s] & $10050\pm 100$&$10030\pm 64$ &$10050\pm 100$ \\ 
$ \tilde\mu $ &$ -18.5829\pm 0.0054$ & $-18.5848\pm 0.0042$ & $-18.5830\pm 0.0053$ \\
\hline
$M_b$ & $-$ & $-19.426\pm 0.011$& $-19.253\pm 0.028$\\
$r_d(z^*)~[{\rm Mpc}]$ &$-$ & $147.72\pm 0.21$& $136.9^{+2.0}_{-2.2}$ \\
\hline

$h$
	 &$-$ & $0.6788\pm 0.0037$& $ 0.7347\pm 0.0096$\\ 
$\omega_m$ &$-$& $0.14109\pm 0.00079$& $0.1681\pm 0.0070 $\\
\hline
$100\omega_b$ & $-$ & $2.224\pm 0.013$ &$2.79\pm 0.15$ \\
$\omega_{\rm cdm}$ &$-$ & $0.11821\pm 0.00083$&$0.1395\pm 0.0068$ \\
$S_8$ & $-$& $0.812\pm 0.011$ & $0.890\pm 0.044$\\
$t_u$ [Gyr] & $-$& $13.804\pm 0.018$& $12.70\pm 0.19$\\
\hline
\end{tabular}
\caption{Mean $\pm 1\sigma$  of reconstructed parameters in the $\Lambda$CDM model in light of  DESI and Pantheon$^+$ data either uncalibrated, calibrated with Planck or with S$H_0$ES.}
\label{tab:cosmoparam}
\end{table*}

\begin{table*}[]
    \centering
    \begin{tabular}{|l|c|c|c|}
     \hline
        \multicolumn{4}{|c|}{$w_0w_a$CDM+$\Omega_k$}\\
        \hline\rule{0pt}{3ex}
        & $\mathcal{DP}^+$ & $\mathcal{DP}^+$+Planck & $\mathcal{DP}^+$+S$H_0$ES\\
        \hline
$ \Omega{}_{m } $ & $  0.285^{+0.029}_{-0.022} $ &
$   0.3154\pm 0.0066 $ &
$   0.295^{+0.019}_{-0.023} $  \\
$ H_0r_d(z^*)$ [km/s]  & $ 9978\pm 110 $ &
$   9906\pm 100 $ &
$   9989\pm 110 $  \\
$ \tilde\mu $&$ -18.569\pm 0.010 $ &  $-18.5669\pm 0.0088$&  $-18.568\pm 0.010 $\\
\hline
$ M_b $ & $ - $ &
$   -19.430\pm 0.019 $ &
$   -19.251\pm 0.026 $  \\
$ r_s(z_d)~[{\rm Mpc}] $ & $ - $ &
$  147.42\pm 0.26 $ &
$   136.9\pm 2.1 $  \\
\hline
$ h $ & $ -$ & 
$  0.6720\pm 0.0068 $ &
$ 0.7300\pm 0.0098 $ \\
$ \omega_m $& $- $ &
$  0.1424\pm 0.0011 $ &
$ 0.157^{+0.011}_{-0.012} $ \\

\hline
$ 100\omega{}_{b } $ & $ - $ &
$   2.219\pm 0.014 $ &
$  3.16\pm 0.40 $  \\
$ \omega{}_{\rm cdm } $ & $ - $ &
$  0.1195\pm 0.0011 $ &
$  0.125^{+0.014}_{-0.016} $  \\
$ S_8 $ & $ - $ &
$   0.838\pm 0.012 $ &
$   0.73^{+0.12}_{-0.17} $  \\
$ t_U $ [Gyr] & $-$ &
$ 13.841\pm 0.070 $ &
$  12.66\pm 0.20 $  \\
\hline
$ w_0 $ & $  -0.92\pm 0.10 $ &
$  -0.839\pm 0.066 $ &
$   -0.906^{+0.11}_{-0.096} $ \\
$ w_a $ & $-0.49^{+0.90}_{-0.67} $ &
$   -0.75^{+0.34}_{-0.29} $ &
$   -0.72^{+0.74}_{-0.58} $ \\
$ \Omega{}_{k }$ & $ 0.0997\pm 0.081 $ &
$   -0.0020\pm 0.0019 $ &
$   0.091\pm 0.076 $ \\
\hline
\end{tabular}
\caption{ Mean $\pm 1\sigma$ of reconstructed parameters in the $w_0w_a$CDM+$\Omega_k$ model in light of  DESI and Pantheon$^+$ data either uncalibrated, calibrated with Planck or with S$H_0$ES.}
\label{tab:cosmoparam1}
\end{table*}

\begin{table*}[]
    \centering
    \begin{tabular}{|l|c|c|c|c|}
     \hline
        & \multicolumn{2}{|c|}{EDE}& \multicolumn{2}{|c|}{varying $m_e$}\\
        \hline\rule{0pt}{3ex}
    late-time    & flat-$\Lambda$CDM & $w_0w_a+\Omega_k$ free &  flat-$\Lambda$CDM & $w_0w_a+\Omega_k$ free\\
          \hline 
          $ \Omega{}_{m } $ & $ 0.2970\pm 0.0045 $ &
$   0.2980\pm 0.0060 $ &
$   0.2857\pm 0.0047 $ &
$   0.2859\pm 0.0052 $ \\
          $ H_0r_d(z^*)~[km/s] $ & $ 10140\pm 62 $ &
$  10150\pm 100 $ &
$  10320\pm 68 $ &
$  10030\pm 100 $ \\
$ \tilde\mu $&$ -18.5879\pm 0.0041 $ &
$ -18.5692\pm 0.0093 $ &
$  -18.5916\pm 0.0041 $ &
$  -18.5725\pm 0.0087 $ \\
\hline
$ M_b $ & $ -19.310\pm 0.023 $ &
$  -19.309\pm 0.022 $ &
$   -19.319\pm 0.020 $ &
$   -19.262^{+0.027}_{-0.024} $ \\
$ r_s(z_d)~[{\rm Mpc}] $ & $ 141.5\pm 1.3 $ &
$  142.8\pm 1.5  $ &
$   144.19\pm 0.73 $ &
$  137.8\pm 2.0 $ \\
\hline

       $ h $ & $ 0.7170\pm 0.0076 $ &
$  0.7114\pm 0.0079 $ &
$  0.7155\pm 0.0070 $ &
$  0.7280^{+0.0096}_{-0.0082} $ \\
$ \omega_m $ & $ 0.1527\pm 0.0029 $ &
$   0.1508\pm 0.0031 $ &
$   0.1462\pm 0.0014 $ &
$   0.1515\pm 0.0024 $ \\

\hline
$ 100\omega{}_{b } $ & $ 2.258\pm 0.019 $ &
$   2.250\pm 0.019 $ &
$   2.259\pm 0.013 $ &
$   2.387\pm 0.037 $ \\
$ \omega{}_{\rm cdm } $ & $ 0.1294\pm 0.0029 $ &
$  0.1276\pm 0.0030$ &
$  0.1230\pm 0.0014 $ &
$  0.1270\pm 0.0021 $ \\
$ S_8$ & $  0.831\pm 0.012 $ &
$    0.837\pm 0.012$ &
$   0.8127\pm 0.0088 $ &
$   0.8089\pm 0.0094 $ \\
$ t_U $ [Gyr] & $ 13.18\pm 0.12 $ &
$   13.27\pm 0.14$ &
$   13.352\pm 0.083 $ &
$  12.91\pm 0.16 $  \\
  \hline 
$ w_0 $  & $-$ & $ -0.834\pm 0.069 $& $-$ &
$ -0.816\pm 0.071$ \\
$w_a$& $-$ & $  -0.76\pm 0.32 $& $-$ &
$ -0.83\pm 0.34 $ \\
$ \Omega{}_{k } $& $-$ & $ 0.0000^{+0.0016}_{-0.0018} $& $-$ &
$ -0.0114\pm 0.0031 $ \\
\hline

$f_{\rm EDE}(z_c) $& $ 0.113^{+0.024}_{-0.020} $ & $  0.084\pm 0.028 $ & $-$ & $-$ \\
$\theta_i$ & $ 2.73^{+0.22}_{+0.011} $ &
$ 	2.42^{+0.62}_{+0.050} $ & $-$ & $-$ \\
$ \log_{10}z_c$ & $3.609^{+0.066}_{-0.13} $ &
$ 3.60\pm 0.15 $ &$-$ & $-$ \\
$ \Delta m_e/m_e $& $-$ & $-$ & $ 0.0213\pm 0.0042 $ & 
$ 0.071\pm 0.015 $ \\
\hline
$Q_{\rm DMAP}$ & $3.1\sigma$& $3.1\sigma$& $3.5\sigma$&$1.1\sigma$ \\
\hline

\end{tabular}
\caption{Mean $\pm 1\sigma$  (or $2\sigma$ for one-sided bounds) of reconstructed parameters in the EDE and varying $m_e$ models in light of  Planck+DESI+Pantheon$^+$+S$H_0$ES, when the late-time dynamics is either fixed to $\Lambda$CDM or let free to vary with $w_0w_a+\Omega_k$.}
\label{tab:cosmoparam2}
\end{table*}

\begin{table*}[]
    \centering
    \begin{tabular}{|l|c|c|c|c|}
     \hline
        & \multicolumn{2}{|c|}{WZDR}& \multicolumn{2}{|c|}{PMF}\\
        \hline\rule{0pt}{3ex}
    late-time    & flat-$\Lambda$CDM & $w_0w_a+\Omega_k$ free &  flat-$\Lambda$CDM & $w_0w_a+\Omega_k$ free\\
          \hline 

$ \Omega{}_{m } $ & $ 0.2920\pm 0.0043 $ &
$ 0.2962\pm 0.0059 $ &
$ 0.2930\pm 0.0044 $ &
$0.2928\pm 0.0055 $ \\

$ H_0r_d(z^*)~[{\rm km/s}] $ & $ 10230\pm 60 $ &
$ 10120\pm 100 $ &
$ 10210\pm 62 $ &
$ 10260\pm 93 $ \\
$ \tilde\mu $ & $ -18.5922\pm 0.0046 $ &
$ -18.5744\pm 0.0096 $ &
$ -18.5892\pm 0.0040 $ &
$ -18.5721\pm 0.0090 $ \\
\hline
$ M_b $ & $ -19.294\pm 0.023 $ &
$ -19.283\pm 0.023 $ &
$ -19.358^{+0.017}_{-0.014} $ &
$ -19.346\pm 0.017 $ \\
$ r_s(z_d)~[{\rm Mpc}] $ & $ 141.3\pm 1.2 $ &
$ 140.3\pm 1.8 $ &
$ 145.55\pm 0.56 $ &
$ 146.52^{+0.69}_{-0.52} $ \\
\hline
$ h $ & $0.7239\pm 0.0080 $ &
$ 0.7215\pm 0.0084 $ &
$ 0.7017^{+0.0055}_{-0.0047} $ &
$ 0.7003\pm 0.0064 $ \\
$ \omega_m $ & $ 0.1530\pm 0.0026 $ &
$ 0.1542\pm 0.0033 $ &
$ 0.1443\pm 0.0012 $ &
$ 0.1436\pm 0.0013 $ \\
$ 100\omega{}_{b } $ & $ 2.249^{+0.013}_{-0.015} $ &
$ 2.260\pm 0.017 $ &
$ 2.237\pm 0.012 $ &
$ 2.223\pm 0.014 $ \\
$ \omega{}_{cdm } $ & $ 0.1299\pm 0.0026 $ &
$  0.1310\pm 0.0033 $ &
$  0.1212\pm 0.0012 $ &
$  0.1207\pm 0.0013 $ \\
$ S_8 $ & $ 0.813\pm 0.010 $ &
$ 0.816\pm 0.011 $ &
$ 0.8118^{+0.0079}_{-0.0089} $ &
$ 0.8223\pm 0.0096 $ \\
$ t_U $ [Gyr] & $ 13.12\pm 0.12 $ &
$ 13.07\pm 0.16 $ &
$ 13.519^{+0.054}_{-0.067} $ &
$ 13.595\pm 0.073 $ \\
\hline

$ w_0 $ & $-$ &  $-0.854\pm 0.064 $ & $-$ & $-0.813\pm 0.069 $\\
$ w_a $ &  $-$ & $-0.55^{+0.31}_{-0.26} $ &$-$ & $-0.97\pm 0.33 $\\
$ \Omega{}_{k } $ &  $-$ & $-0.0032\pm 0.0019 $ & $-$& $-0.0001^{+0.0013}_{-0.0016}$\\
\hline
$ \Delta N_{\rm ur} $ & $ 0.77\pm 0.15 $ & $ 0.96\pm 0.26 $ &  $-$  &  $-$  \\
$ \log_{10}(z_t)$ & $ 4.06^{+0.28}_{-0.098} $ & $3.96^{+0.32}_{-0.15} $ &  $-$  &  $-$  \\
$b_{\rm clump}$ & $-$ & $-$ & $ 0.48\pm 0.14 $ &
$0.191^{+0.068}_{-0.17} $  \\ 
\hline
$Q_{\rm DMAP}$ & $2.3\sigma$& $2.1\sigma$& $5.2\sigma$& $4.7\sigma$ \\
\hline

\end{tabular}
\caption{Mean $\pm 1\sigma$  (or $2\sigma$ for one-sided bounds) of reconstructed parameters in the WZDR and PMF models in light of  Planck+DESI+Pantheon$^+$+S$H_0$ES, when the late-time dynamics is either fixed to $\Lambda$CDM or let free to vary with $w_0w_a+\Omega_k$.}
\label{tab:cosmoparam3}
\end{table*}

\section{Comparison with Gaussian Processes}
\label{app:GP}

\begin{figure*}
    \centering
    \includegraphics[width=2\columnwidth]{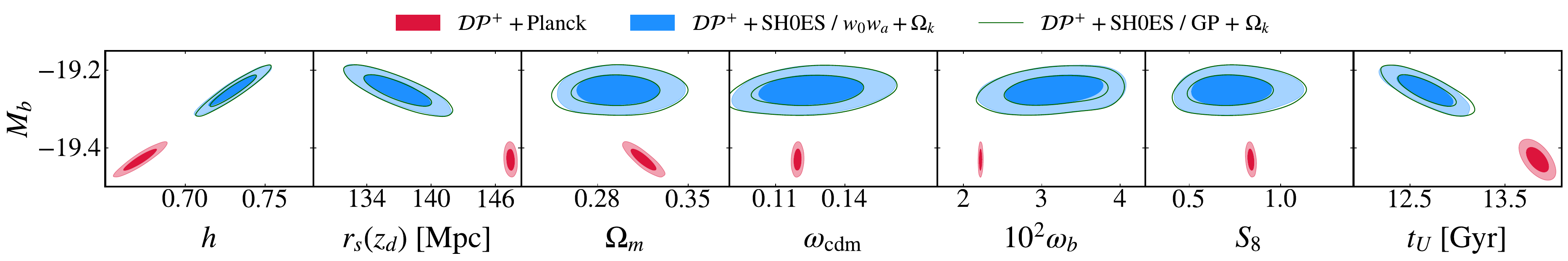}
    \caption{A comparison between constraints in the $w_0w_a+\Omega_k$ model and the Gaussian Process reconstruction of $w(z)$ with $\Omega_k$ free. The  CPL parametrisation gives enough freedom to capture deviations from $\Lambda$ at late-times.}
    \label{fig:GP}
\end{figure*}

In this appendix, we show that considering the CPL parameterization of the DE equation of state, $w(a)=w_0 + w_a(1-a)$, together with a curved universe, quantified by $\Omega_k$, is flexible enough to capture the information about the expansion history contained in the BAO and SN1a data considered in this work. To do so, we perform a model-independent reconstruction of the equation of state of DE using a Gaussian Process (GP) along the lines of \cite{Calderon:2022cfj,Hwang:2022hla}
\begin{equation}\label{eq:w_GP}
    w(a)\sim \mathcal{GP}(\bar f=w_0+w_a(1-a),K=k(\sigma_f,\ell_f))~,
\end{equation}
where $\bar f$ denotes the mean function, which we take to be of the CPL form, and $k(\sigma_f,\ell_f)$ is an exponential-squared kernel given by 
\begin{equation}
    k(\sigma_f,\ell_f)=\sigma_f^2~\exp{\left[-\frac{(x-x')^2}{2\ell_f^2}\right]}~.
\end{equation}
This is implemented in a modified version of \texttt{class}. In practice, we marginalize over the space of mean functions by varying $(w_0,w_a)$, together with the hyperparameters $(\sigma_f,\ell_f)$ describing the extra-correlations (or deviations away from $w_0w_a\rm CDM$). Such form in \eqref{eq:w_GP} gives us enough freedom to marginalize over a wide range of expansion histories in a ``non-parametric'' manner, i.e. without restricting ourselves to linear functions of $a$.
We show our results in Fig.~\ref{fig:GP}, explicitly confirming that the Gaussian Process modelling of $w(a)$ yields essentially the same constraints\footnote{It has also been shown that the combined data does not suggest more complicated dynamics, beyond $w_0$-$w_a$  \cite{DESI:2024aqx}.} as the $w_0w_a{\rm CDM}+\Omega_k$ presented in the main text. As the latter parametrization yields much faster MCMC convergence and is easier to interpret and reproduce, we focus on it in the main text when presenting our results.

\section{Results for WZDR and PMF}
\label{app:WZDR_PMF}
\subsection{PMF}
\label{app:PMF}
\begin{figure*}
    \centering
    \includegraphics[width=2\columnwidth]{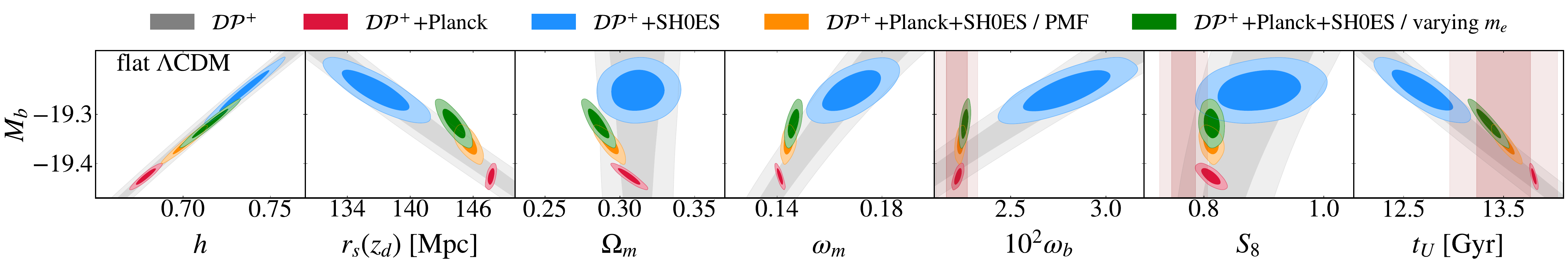}
    \includegraphics[width=2\columnwidth]{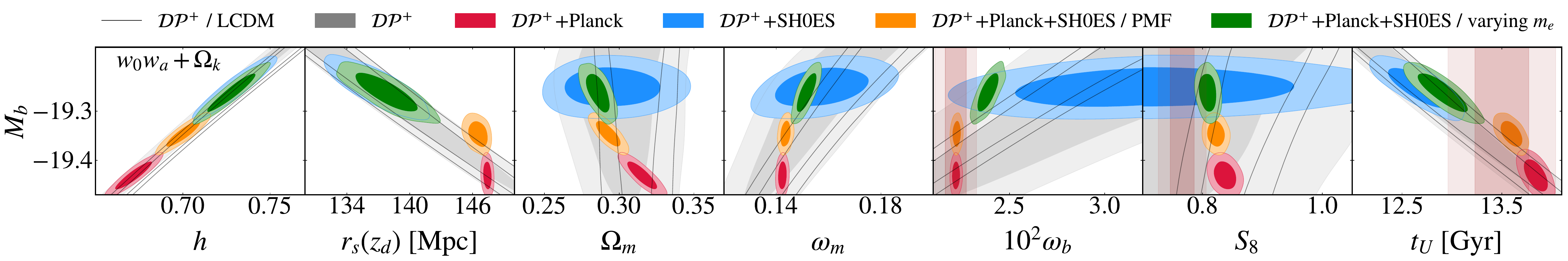}
     \caption{Same as Fig.~\ref{fig:LCDM_EDE_me} in the main text, now comparing the PMF and varying $m_e$ models reconstructed from a fit to ${\cal DP}^+$+Planck+S$H_0$ES, assuming $\Lambda$CDM at late-times (top panel) or  allowing for freedom in the behavior of Dark Energy and curvature at late-times (bottom panel).}
    \label{fig:LCDM_PMF_me}
\end{figure*}

We model primordial magnetic fields following the 3-zone model from Ref.~\cite{Jedamzik:2020krr} using a publicly available modified version of the \texttt{CLASS} code.\footnote{\url{https://github.com/GuillermoFrancoAbellan/class_clumpy}} We let the clumping factor $b_{\rm clump}$ free to vary in a large flat prior. We perform the same set of MCMC analyses as for the EDE and varying $m_e$ model, analyzing the model in light of  ${\cal DP}^+$+Planck+S$H_0$ES, assuming either that $\Lambda$CDM holds at late-times or  allowing for freedom in the behavior of Dark Energy and curvature. Our results are provided in  Tab.~\ref{tab:cosmoparam3} and shown in Fig.~\ref{fig:LCDM_PMF_me}, compared with that of the varying $m_e$ model. One can see that the PMF model follows the same degeneracy directions as the varying $m_e$ model although it performs less well. In fact, we find residuals tensions of $\sim 5.2\sigma$ when assuming flat $\Lambda$CDM at late-times and $\sim 4.7$ when leaving $w_0w_a+\Omega_k$ free to vary. This is much worse than the varying $m_e$ model. It is clear that the impact of $m_e$ on the Thomson scattering cross-section plays a big role in this result, as discussed in the main text. Finally, we show in Fig.~\ref{fig:CPL_Omk_WZDR_PMF} the reconstructed posteriors of $\{h,w_0,w_a,\Omega_k\}-$vs$-M_b$. The preference for dynamical dark energy remains in this model, and the universe is compatible with being flat contrarily to the varying $m_e$ model. More work is required to fully understand the reason for these differences. 

\subsection{WZDR}
\label{app:WZDR}

The Wess-Zumino dark radiation (WZDR) model (also called `stepped dark radiation') \cite{Aloni:2021eaq} introduces a self-interacting dark radiation component that appears after BBN. The self-interactions are mediated by a massive scalar field with a mass around the eV scale. When the mediator becomes non-relativistic it annihilates into additional dark radiation, causing an increase (i.e., `step') in the dark radiation energy density. The model is described by two parameters: the initial dark radiation energy density (quantified by the number of additional effective neutrinos when the scalar mediator is ultra-relativistic, $\Delta N_{\rm ur}$) and the redshift at which the scalar mediator transitions to being non-relativistic, $z_t$. Regarding $\Delta N_{\rm ur}$, we focus on the region of parameter space $\Delta N_{\rm ur}>0$  and make use of a large flat prior otherwise. For $z_t$ we use a logarithmic prior $\log_{10}(z_t)\in[3,5]$.
As previously, we run MCMC analyses analyzing the model in light of  ${\cal DP}^+$+Planck+S$H_0$ES, assuming either that $\Lambda$CDM holds at late-times or  leaving $w_0w_a+\Omega_k$ free to vary. Our results are provided in Tab.~\ref{tab:cosmoparam3} and shown in Fig.~\ref{fig:CPL_Omk_WZDR_PMF}, compared with that of the EDE model. One can see that the WZDR model follows the same degeneracy directions as the EDE model, and in fact performs better. We find residuals tensions of $\sim 2.3\sigma$ when assuming flat $\Lambda$CDM at late-times and $\sim 2.1$ when leaving $w_0w_a+\Omega_k$ free to vary. This is better than EDE, but worse than the varying $m_e$ model with $w_0w_a+\Omega_k$ free to vary. However, there is no BBN tension in that model, and $S_8$ is also reduced, closer to the values measured from weak lensing observations. It is clear that the specific details of the expansion history, and the dynamics of perturbations associated with the interacting DR, play an important role in the success of the model. Additional insights about the model success, and the role of the step, can be found in Refs~\cite{Aloni:2021eaq,Schoneberg:2023rnx}. 
   Finally, we show in Fig.~\ref{fig:CPL_Omk_WZDR_PMF} the reconstructed posteriors of $\{h,w_0,w_a,\Omega_k\}-$vs$-M_b$. The preference for dynamical dark energy remains in this model (but the universe is compatible with being flat).
\begin{figure*}
    \centering
    \includegraphics[width=2\columnwidth]{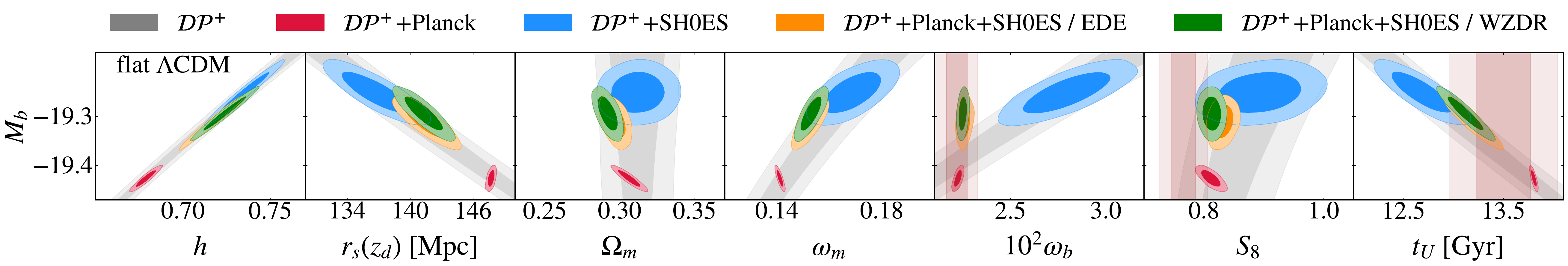}
    \includegraphics[width=2\columnwidth]{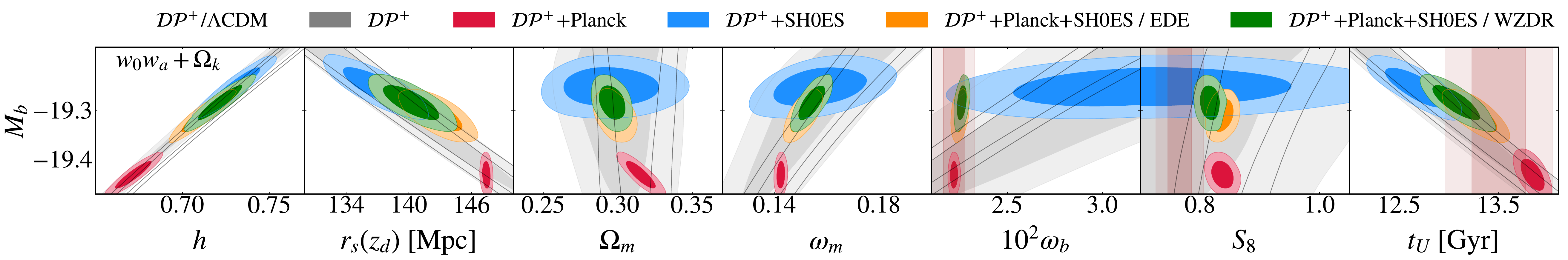}

    \caption{Same as Fig.~\ref{fig:LCDM_EDE_me} in the main text, now comparing the EDE and WZDR models reconstructed from a fit to ${\cal DP}^+$+Planck+S$H_0$ES, assuming $\Lambda$CDM at late-times (top panel) or  allowing for freedom in the behavior of Dark Energy and curvature at late-times (bottom panel).}
    \label{fig:LCDM_EDE_WZDR}
\end{figure*}

\begin{figure*}
    \centering
    \includegraphics[width=2\columnwidth]{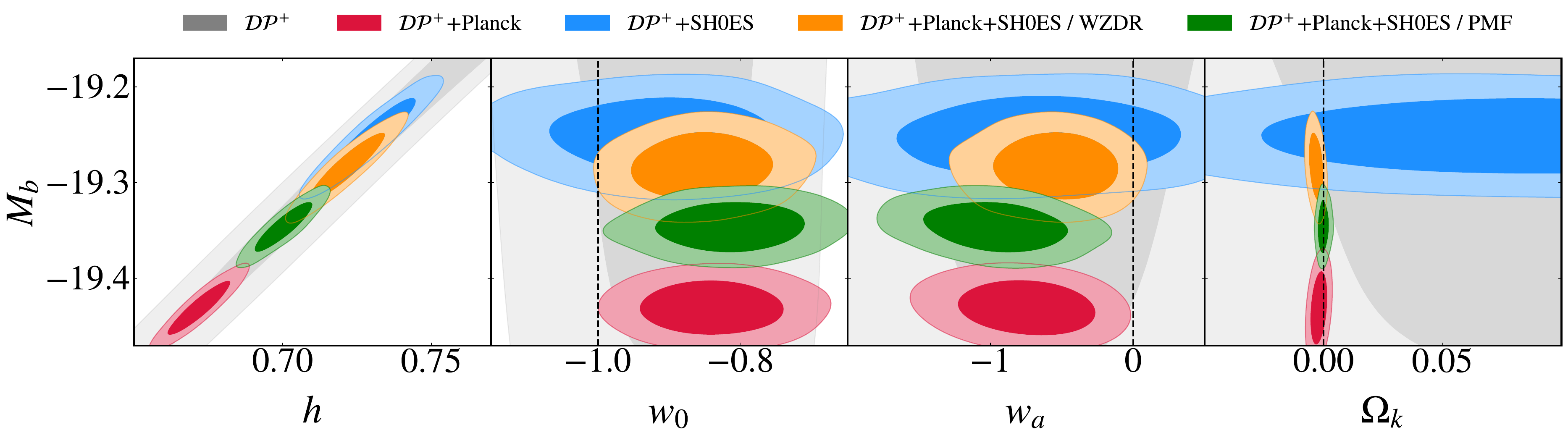 }

     \caption{Same as Fig.~\ref{fig:CPL_Omk} in the main text, now comparing the PMF and WZDR models reconstructed from a fit to ${\cal DP}^+$+Planck+S$H_0$ES.}
    \label{fig:CPL_Omk_WZDR_PMF}
\end{figure*}

\begin{figure*}
    \centering
    \includegraphics[width=2\columnwidth]{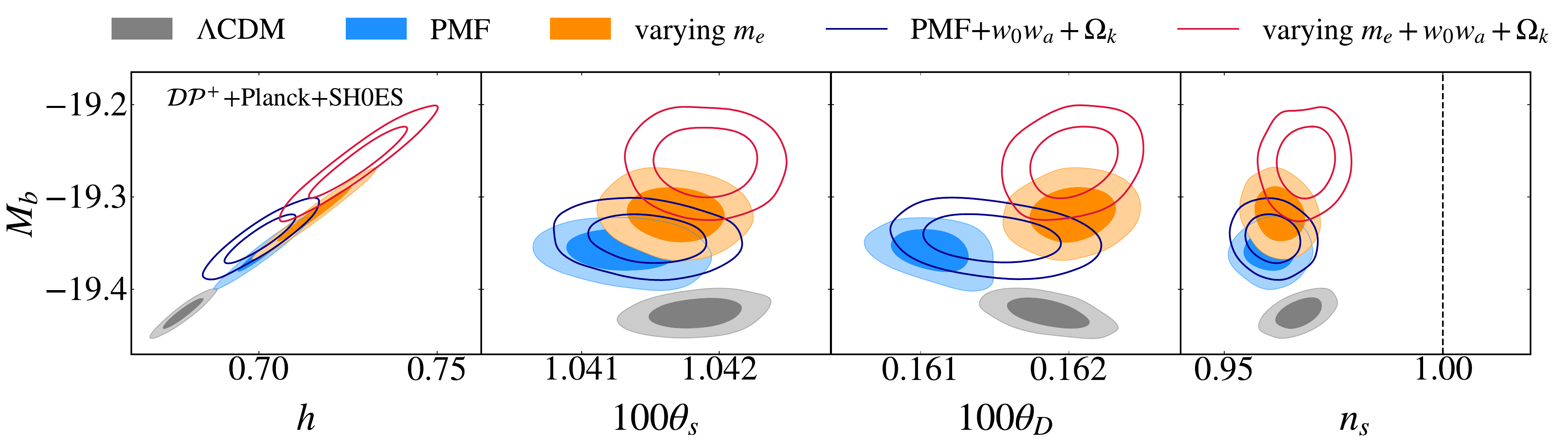}
    \includegraphics[width=2\columnwidth]{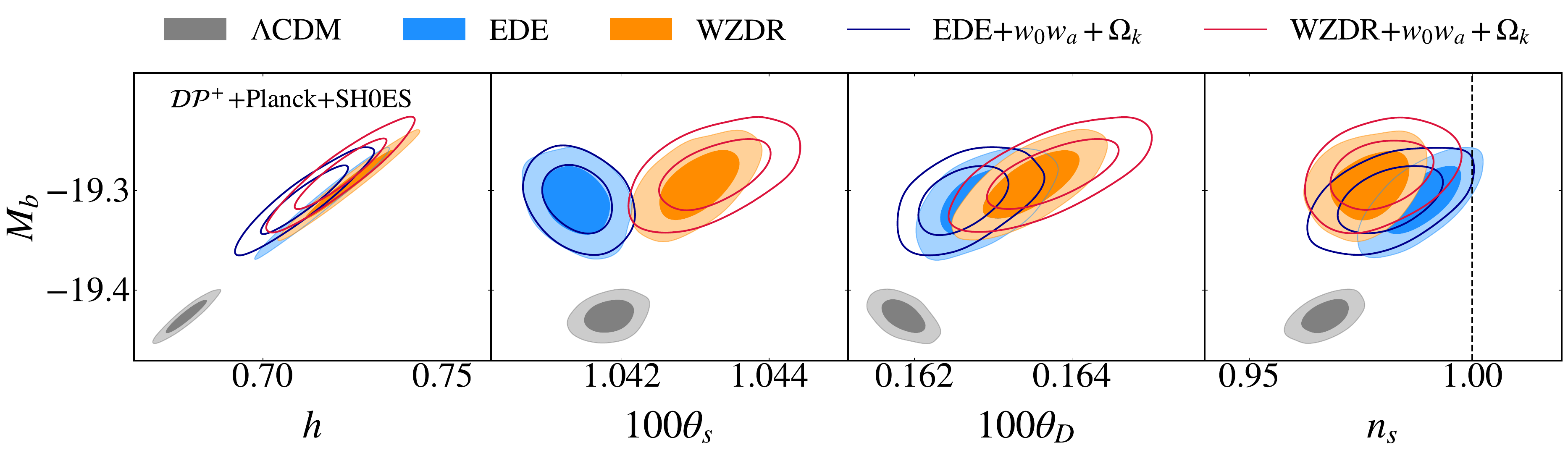}

    \caption{Same as Fig.~\ref{fig:thetaD_ns} in the main text, now comparing the PMF and $m_e$ models (top panel) and the EDE and WZDR models (bottom panel) reconstructed from a fit to ${\cal DP}^+$+Planck+S$H_0$ES.}
    \label{fig:thetaD_ns_WZDR}
\end{figure*}

\bibliography{biblio}% Produces the bibliography via BibTeX.
\end{document}